\documentclass[twocolumn, appendixfloats, revtex4, numberedappendix, iop]{openjournal}

\usepackage{CJK}

\usepackage{enumitem}
\usepackage{url}

\usepackage[most,breakable]{tcolorbox}

\shorttitle{Attenuation Bias in Astronomical Data-Driven Models}
\shortauthors{Ting}

\begin{document}
\begin{CJK*}{UTF8}{gbsn}

\title{Why Machine Learning Models Systematically Underestimate Extreme Values \vspace{-1.5cm}}

\author{Yuan-Sen Ting (丁源森)}
\affiliation{Department of Astronomy, The Ohio State University, Columbus, OH 43210, USA}
\affiliation{Center for Cosmology and AstroParticle Physics (CCAPP), The Ohio State University, Columbus, OH 43210, USA}

\begin{abstract}
    A persistent challenge in astronomical machine learning is a systematic bias where predictions compress the dynamic range of true values---high values are consistently predicted too low while low values are predicted too high. Understanding this bias has important consequences for  astronomical measurements and our understanding of physical processes in astronomical inference. Through analytical examination of linear regression, we show that this bias arises naturally from measurement uncertainties in input features and persists regardless of training sample size, label accuracy, or parameter distribution. In the univariate case, we demonstrate that attenuation becomes important when the ratio of intrinsic signal range to measurement uncertainty ($\sigma_{\text{range}}/\sigma_x$) is below $\mathcal{O}(10)$---a regime common in astronomy. We further extend the theoretical framework to multivariate linear regression and demonstrate its implications using stellar spectroscopy as a case study. Even under optimal conditions---high-resolution APOGEE-like spectra ($R=24,000$) with high signal-to-noise ratios (SNR=100) and multiple correlated features---we find percent-level bias. The effect becomes even more severe for modern-day low-resolution surveys like LAMOST and DESI due to the lower SNR and resolution. These findings have broad implications, providing a theoretical framework for understanding and addressing this limitation in astronomical data analysis with machine learning.
\end{abstract}

\section{Introduction}

The advent of modern astronomical surveys has revolutionized our understanding of the universe. Large-scale observational programs have amassed unprecedented volumes of high-quality data, enabling detailed studies across multiple astronomical domains \citep{Ivezic2012,Weinberg2013,LSST2017,Aerts2021}. This wealth of information has motivated the development of data-driven models and machine learning techniques to extract meaningful insights from complex astronomical phenomena \citep{Baron2019,Huertas-Company2023,Smith2023}.

Given that many physical simulated models continue to suffer from systematic errors, data-driven approaches have become increasingly popular for astronomical inference problems \citep{Casey2016,Olney2020,Green2021,Huang2022,Lin2022}. These approaches leverage both observational data and more reliable target properties from other sources to establish relationships between them. For example, in stellar spectroscopy, data-driven models have emerged as powerful tools for inferring fundamental properties such as stellar ages and distances, where `golden' age measurements come from asteroseismology and precise distances from parallax measurements of nearby stars  \citep{Ness2018,Casali2019,Hogg2019}.

Even for more routine applications, such as determining basic stellar properties, data-driven methods have proven valuable \citep{Fabbro2018,Leung2019,Anders2023}. This is particularly true for low-resolution spectra, where theoretical spectral models often fall short. A common approach is to transfer stellar parameters and elemental abundances measured from high-resolution surveys like APOGEE to lower-resolution surveys such as RAVE \citep{Casey2017,Guiglion2020}, LAMOST \citep{Ho2017,Ting2017,Xiang2019,Zhang2020,Li2022}, DESI \citep{Zhang2024}, or Gaia \citep{Angelo2024,Hattori2024,Khalatyan2024,Li2024}. These low-resolution observations, especially of distant stars or those obtained as part of larger cosmological surveys, typically achieve lower  spectral resolution and signal-to-noise ratios (SNR).

In astronomical data analysis, mapping between observables (e.g., spectra) and physical parameters (e.g., stellar properties) presents no inherent distinction between independent and dependent variables. Two main approaches exist in this field. The first approach uses generative models (e.g., The Payne, The Cannon, Cycle-StarNet) that map labels to observables and use $\chi^2$ minimization or MCMC sampling to obtain labels \citep{Dafonte2016,Rix2016,Ting2016,Chandra2020,Straumit2022,Purmortal2022,Laroche2024,Ting2019,Rozanski2024,Ness2015,OBriain2021}. The second approach employs discriminative models (e.g., AstroNN, StarNet) that directly map observables to labels \citep{Bailer-Jones1997,Snider1999,Bialek2020,Garraffo2021,Ambrosch2023,Gabran2023,Zhang2023,Fallows2024,Leung2019,Fabbro2018}.

\begin{figure*}[htbp]
    \centering
    \includegraphics[width=\textwidth]{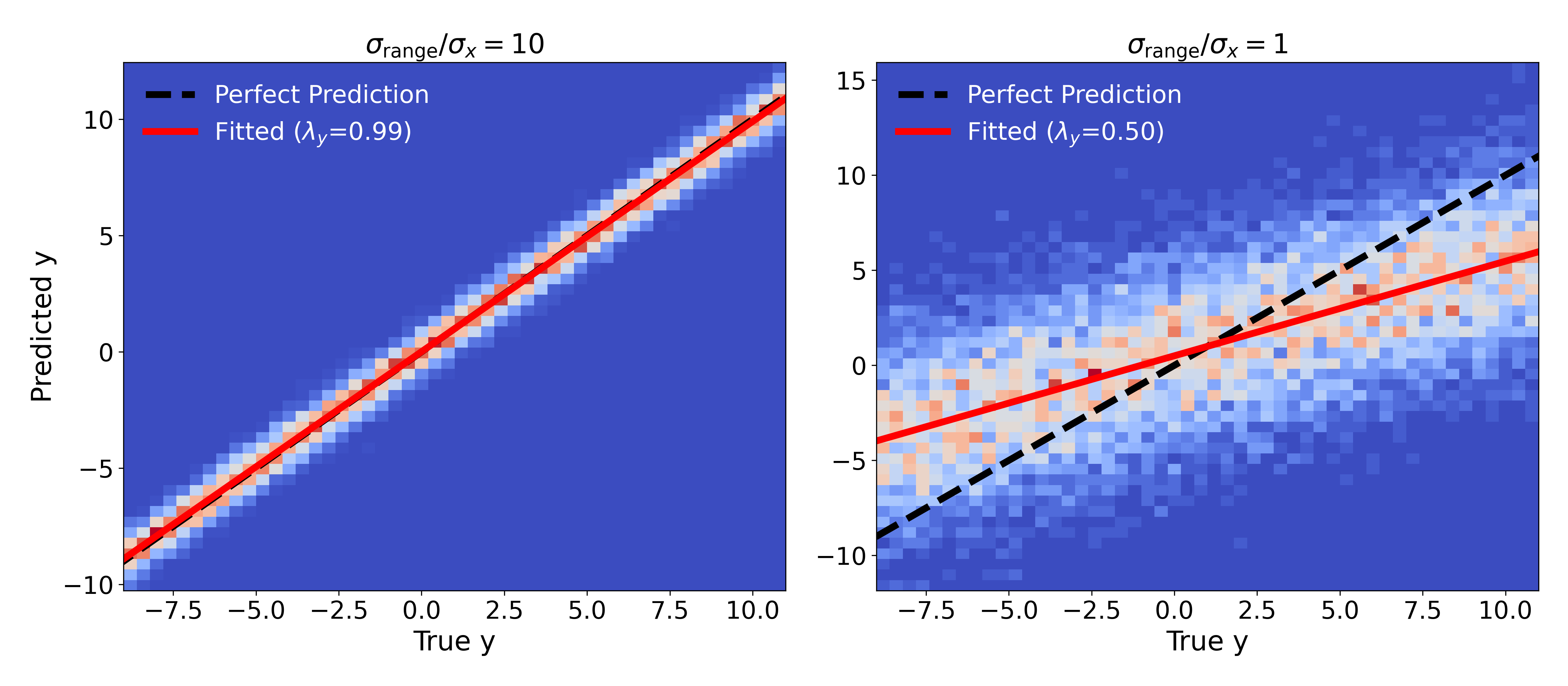}
    \caption{Visualization of attenuation bias through density plots comparing predicted versus true values. Left panel shows predictions with $\sigma_{\text{range}}/\sigma_x = 10$, where the fitted regression line between the predicted $y$ as a function of true $y$ (red solid, $\lambda_y = 0.99$) closely follows the perfect prediction line (black dashed) with only 1\% bias. Right panel shows the case where measurement uncertainties are comparable to signal variance ($\sigma_{\text{range}}/\sigma_x = 1$), where the fitted line ($\lambda_y = 0.5$).}
    \label{fig:s_shaped_bias}
 \end{figure*}

A persistent challenge in data-driven models is systematic bias in the predictions compared to ground truth. This bias typically manifests as predictions being too low at high values and too high at low values (e.g., see the right panel of Figure~\ref{fig:s_shaped_bias}). In this study, we aim to demonstrate that this bias is not due to model training or sample distribution issues, as it persists even with uniformly distributed training data in both feature and label spaces with perfect ground truth labels. We show that the bias depends only on the input measurement uncertainties, which systematically attenuate the regression coefficients and, consequently, the inferred label.

In linear regression models, this systematic effect, known as attenuation bias or regression dilution, has been studied analytically. \citet{Fuller1987} provided a comprehensive mathematical treatment of measurement error models, establishing the fundamental relationship for univariate cases. Subsequent work by \citet{MacMahon1990} and \citet{Rosner1990} extended these concepts to epidemiological applications, while \citet{Carroll1995} developed methods for nonlinear measurement error models. \citet{Frost2002} provided accessible reviews for practitioners. In astronomy, \citet{Kelly2007,Kelly2012} and \citet{Loredo2004} developed methods to handle measurement errors and source uncertainties in survey analysis, though these works primarily examined univariate relationships or treated multivariate cases without explicitly analyzing the role of feature correlations in mitigating bias.

The emergence of modern machine learning methods and high-dimensional data analysis presents new challenges that warrant revisiting these fundamental insights. Specifically, our work extends the classical framework in several ways: First, we examine the low signal-to-noise regime, which, while extreme for epidemiological studies, is typical in astronomical applications. Second, we demonstrate how attenuation bias scales also applies to nonlinear transformations, which is relevant for modern machine learning architectures. Third, we provide analytical treatment for multivariate cases with correlated features—a scenario largely unexplored in both classical and astronomical literature but ubiquitous in spectroscopic analysis where physical parameters generate correlated spectral features. Finally, we provide the first systematic examination of these effects in the context of astronomical machine learning, with particular focus on stellar spectroscopy, where the combination of high dimensionality (hundreds to thousands of spectral pixels), significant measurement uncertainties, and naturally correlated features creates a unique statistical regime that directly impacts modern data-driven approaches to stellar parameter estimation.

Our focus on linear regression stems from several considerations. First, many practical astronomical applications effectively employ linear models—for example, The Cannon, a widely-used tool in stellar spectroscopy, is fundamentally a linear regression model with polynomial features. Second, the mathematical simplicity of linear models enables us to derive concrete, analytical solutions that clearly demonstrate these effects. As many regression tasks, including Gaussian Process and neural networks, can be reframed as generalized linear regression, this framework provides insights into the broader challenges of using machine learning for direct label inference through data-driven models.

The paper is organized as follows: Section~\ref{sec:theory} provides a theoretical framework and mathematical derivations of attenuation bias, including both univariate and multivariate cases. Section~\ref{sec:spectra} explores our empirical analysis using spectra and the implications for spectra-to-label models. Section~\ref{sec:discussion} broadens the discussion to the impact of attenuation bias, providing key considerations to mitigate and calibrate such bias, and its effects on fundamental relations, while Section~\ref{sec:conclusion} summarizes our findings.

\section{Attenuation Bias}
\label{sec:theory}

In astronomical analyses involving mappings between observables and physical parameters, we adopt the convention of treating observables (e.g., spectra) as input features $x$ and physical parameters as target labels $y$. We focus on direct inference using error-ignorant discriminative models—where measurement uncertainties in input features $x$ are ignored during model training and inference. While this simplifies the full statistical problem, it reflects standard practice in machine learning applications due to the practical challenges of error quantification and incorporation in model architectures, especially in modern-day machine learning.

Attenuation bias, also known as regression dilution or errors-in-variables bias, systematically underestimates regression coefficients when independent variables contain measurement errors \citep{Fuller1987,Carroll1995}. While classical regression assumes error-free inputs with all uncertainty contained in the dependent variable, real astronomical observations invariably include measurement uncertainties in both dependent and independent variables. These systematic and statistical uncertainties can arise from multiple sources, including instrumental noise from thermal effects and detector imperfections, photon statistics, calibration errors in wavelength and flux measurements, atmospheric distortions in ground-based observations \citep{Stubbs2006,Stubbs2007,Burke2010,Noll2012,Li2016}, and uncertainties in derived stellar parameters used as training labels.

To develop an understanding of this effect, we first examine the mathematical framework in the univariate case that describes how measurement errors propagate through regression analyses. This framework will provide the foundation for understanding both the causes and consequences of attenuation bias in more the complex multivariate scenarios.

\subsection{Theoretical Framework -- Univariate Linear Regression}
\label{sec:theory-univariate}

Consider a linear relationship between a single dependent variable $y$ and independent variable $x$. By extending the feature vector with a constant term, we can absorb the intercept into the coefficient and write:
\begin{equation}
    y_{\text true} = \beta x_{\text true},
    \label{eq:true_model}
\end{equation}
\begin{equation}
    y_{\text obs} = y_{\text true} + \delta_y,
\end{equation}
where $\beta$ is the slope and $\delta_y$ represents both the measurement uncertainty and intrinsic scatter in $y$ with:
\begin{equation}
    \mathbb{E}[\delta_y] = 0 \text{ and } \operatorname{Var}(\delta_y) = \sigma_y^2.
\end{equation}
The observed input $x_{\text{obs}}$ differs from the true value $x_{\text{true}}$ by unbiased measurement error $\delta_x$:
\begin{equation}
    x_{\text{obs}} = x_{\text{true}} + \delta_x,
\end{equation}
\noindent
where $\delta_x$ is assumed to be independent of both the ground truth $x_{\text{true}}$ and the uncertainty in $y$ ($\delta_y$), with:
\begin{equation}
    \mathbb{E}[\delta_x] = 0 \text{ and } \operatorname{Var}(\delta_x) = \sigma_x^2.
\end{equation}
This independence assumption is reasonable for astronomical applications where observed quantities in both $x$ and $y$ are typically measured independently.

For simplicity, we apply the ordinary least squares (OLS) estimator when regressing $y_{\text obs}$ on $x_{\text{obs}}$, though we note that using Generalized Least Squares (GLS) to properly weight both uncertainties would not materially change our conclusions (see Appendix~\ref{app:gls}). The OLS estimator has an analytic solution for its expected value:
\begin{equation}
    \mathbb{E}[\hat{\beta}] = \frac{\operatorname{Cov}(x_{\text{obs}}, y_{\text obs})}{\operatorname{Var}(x_{\text{obs}})} 
\end{equation}

To understand the bias of $\hat{\beta}$ compared to the true $\beta$, we first examine the numerator -- the covariance between $x_{\text{obs}}$ and $y_{\text obs}$:
\begin{align}
    \operatorname{Cov}(x_{\text{obs}}, y_{\text obs}) &= \operatorname{Cov}(x_{\text{true}} + \delta_x, \beta x_{\text{true}} + \delta_y) \nonumber \\
    &= \beta \operatorname{Cov}(x_{\text{true}} + \delta_x, x_{\text{true}}) \nonumber \\
    & \qquad +  \operatorname{Cov}(x_{\text{true}} + \delta_x, \delta_y).
    \label{eq:covariance}
\end{align}

Given our assumptions about independence between measurement uncertainties and true values, this simplifies to:
\begin{equation}
    \operatorname{Cov}(x_{\text{obs}}, y_{\text{obs}}) = \beta \operatorname{Var}(x_{\text{true}}) \equiv \beta \sigma_{\text{range}}^2,
    \label{eq:covariance_final}
\end{equation}
where $\sigma_{\text{range}}^2$ represents the signal variance (i.e., the ``true spread") of $x$. 

As for the denominator, the variance of observed $x$ is:
\begin{align}
    \operatorname{Var}(x_{\text{obs}}) &= \operatorname{Var}(x_{\text{true}} + \delta_x) \nonumber \\
    &= \sigma_{\text{range}}^2 + \sigma_x^2,
    \label{eq:variance}
\end{align}
again drawing from the assumption that $\delta_x$ is independent of $x_{\text{true}}$.  Combining these results yields:
\begin{equation}
    \mathbb{E}[\hat{\beta}] = \beta \frac{\sigma_{\text{range}}^2}{\sigma_{\text{range}}^2 + \sigma_x^2} = \beta \frac{1}{1 + (\sigma_x/\sigma_{\text{range}})^2}.
\end{equation}

If we define the attenuation factor $\lambda_\beta$:
\begin{equation}
    \lambda_\beta \equiv \frac{1}{1 + (\sigma_x/\sigma_{\text{range}})^2},
    \label{eq:lambda}
\end{equation}
our expression for the expected value of $\hat{\beta}$ becomes:
\begin{equation}
    \mathbb{E}[\hat{\beta}] = \lambda_\beta \beta.
    \label{eq:expected_beta_hat}
\end{equation}

This result reveals that the regression coefficient is systematically biased whenever $x$ has measurement uncertainty, with the attenuation factor $\lambda_\beta$ determined by the ratio of measurement uncertainty to the true spread of the variable ($\sigma_x/\sigma_{\text{range}}$). Since $\lambda_\beta$ is always less than unity, the estimated coefficient is inevitably attenuated toward zero, with the effect becoming more severe as measurement uncertainties increase relative to the true spread of values.

The mechanism underlying this attenuation bias can be understood geometrically: measurement uncertainties in the independent variable $x$ systematically increase the horizontal dispersion of data points. When these uncertainties are ignored in the regression analysis, the increased scatter in $x$ leads to a systematic underestimation of the true slope coefficient. This effect becomes particularly pronounced when the measurement uncertainty is comparable to the range of $x$ values in the dataset.

An insight from the mathematical framework is that measurement uncertainties in the dependent variable $y$ do not contribute to the attenuation bias. While uncertainties in $x$ systematically bias the slope estimate toward zero through the attenuation factor $\lambda_\beta$, uncertainties in $y$ do not affect the expected value of $\hat{\beta}$. This asymmetry stems from the fundamental formulation of ordinary least squares regression, where $x$ and $y$ play mathematically distinct roles in the estimation process. Further, this bias is completely independent of training sample size $n$. No amount of additional data can resolve the bias. The attenuation factor $\lambda_\beta$ depends solely on the ratio of signal variances and the measurement uncertainty of $x$, making it immune to increases in the training sample size.

Recall that, in machine learning applications, our goal is to train a model on a training dataset and then use it to make predictions on new, unseen data. In astronomy, this typically involves learning relationships from a training set with known physical parameters and applying the trained model to make predictions for new observations. In the context of linear regression presented above, the linear coefficient plays the role of the ``machine" (like weights in a neuron in a neural network). 

To elaborate on this connection: just as neural networks learn by adjusting their weights to minimize prediction error, linear regression learns by finding the optimal coefficient that minimizes the squared error between predictions and true values. The coefficient, like neural network weights, is the parameter that encodes what the model has learned from the training data and determines how the model will transform inputs into predictions. However, since we have established that the coefficient is biased, even after successfully training a model to minimize empirical error on the training set, predictions on new data will systematically deviate from the true values for $y$.

To see this mathematically, consider that for a trained linear model where $y_{\text{pred}} = \hat{\beta}x_{\text{obs}} $, and $\mathbb{E}[\hat{\beta}] = \lambda_\beta\beta$, we can examine the expected predictions on new observations. For any new observed value $x_{\text{obs}}$, the expected value of the prediction is:
\begin{align}
    \mathbb{E}[y_{\text{pred}}|x_{\text{obs}}] &= \mathbb{E}[\hat{\beta}x_{\text{obs}}] \nonumber \\
    &= \mathbb{E}[\hat{\beta}]\mathbb{E}[x_{\text{obs}}] \nonumber \\
    &= \lambda_\beta\beta x_{\text{true}} \nonumber \\
    &= \lambda_\beta y_{\text{true}} \nonumber \\
    &\equiv \lambda_y y_{\text{true}}
\end{align}
The separation of expectations in line 2 follows from the independence between $\hat{\beta}$ (determined from the training set) and $x_{\text{obs}}$ (from new observations), while the transition to line 3 uses our assumption of unbiased measurement errors for $x$. This demonstrates that the attenuation factor $\lambda_\beta$ directly translates to a bias in the predicted values for the label $y$, which we can denote as $\lambda_y = \lambda_\beta$. At least in this univariate case, the predictions are systematically biased by the same factor $\lambda$, and the two lambda factors can therefore be used interchangeably.

To understand the practical implications for astronomical applications, consider the case of inferring stellar properties from spectral features. In this univariate case (representing a single spectral feature), we have established that the bias depends solely on the ratio between the standard deviation of the true signal and the measurement error, $\sigma_{\text{range}}/\sigma_x$.

For a concrete example, consider a high-resolution spectroscopic survey with resolution $R \equiv \lambda/\Delta\lambda = 24,000$, where $\lambda$ is the wavelength and $\Delta\lambda$ is the wavelength resolution. Here, $x$ represents the normalized flux values in stellar spectra (i.e., spectra that have been divided by their continuum level to have a mean of approximately 1). In normalized spectra, the intrinsic variation in flux typically spans at most 10\% or $\sigma_{\text{range}} = 0.1$ (see also in Section 3). Even with a high signal-to-noise ratio of SNR = 100 ($\sigma_x = 0.01$), yielding $\sigma_{\text{range}}/\sigma_x = 10$, Equation \ref{eq:lambda} predicts $\lambda_\beta = \lambda_y = 0.99$, corresponding to a 1\% bias. For more typical observations, especially of distant targets in surveys like DESI and LAMOST where SNR = 10 ($\sigma_x = 0.1$), the bias becomes severe with $\lambda_\beta = \lambda_y  = 0.5$, indicating a 50\% underestimation. Even at moderate SNR = 30, we still find $\lambda_\beta = \lambda_y  = 0.9$, implying a 10\% bias. These calculations demonstrate that attenuation bias is non-negligible in spectroscopic machine learning applications even under optimistic conditions, motivating a deeper investigation of this effect.

Finally, for completeness, while measurement uncertainties in $y$ do not affect the attenuation bias in $\hat{\beta}$, they do influence its variance. If both $y$ and $x$ are centered (mean substracted), we can show (see Appendix~\ref{app:variance}) that the variance of the OLS estimator follows:
\begin{equation}
    \operatorname{Var}(\hat{\beta}) = \frac{\sigma_y^2 + \beta^2 \sigma_x^2}{n(\sigma_{\text{range}}^2 + \sigma_x^2)}.
\end{equation}
This shows that, while the bias depends solely on the ratio between measurement uncertainty and signal variance in $x$, the precision of our estimate is affected by measurement uncertainties in both variables $x$ and $y$ and scales with sample size $n$. Thus, while increasing the training set size cannot mitigate the fundamental attenuation bias, it can reduce the statistical uncertainty in our biased estimates.

\subsection{Numerical Demonstrations -- Univariate Linear Regression}

To illustrate our theoretical analysis of attenuation bias in the univariate case, we conducted numerical simulations examining its relationship with measurement uncertainties. We generated synthetic data using a true slope $\beta = 2.0$ and sampled $x_{\text{true}}$ uniformly from $[-5, 5]$. We then added Gaussian measurement errors $\delta_x \sim \mathcal{N}(0, \sigma^2_{x})$ to create $x_{\text{obs}}$. The choice of a uniform distribution for $x_{\text{true}}$ was deliberate to demonstrate that attenuation bias persists even with a balanced, unbiased sample, with no uncertainty in the training label $y$ and a straightforward one-dimensional relationship. This confirms that the bias stems solely from measurement uncertainties in the independent variable rather than from sample imbalance or complex variable interactions.

Although our results show the bias is independent of sample size, we simulated 10,000 samples to better visualize the effect at different measurement uncertainty ratios. Unless stated otherwise, we adopt an 8:2 split for training and validation throughout this study. To generate the predictions shown in Figure~\ref{fig:s_shaped_bias}, we first fitted a linear regression using the observed training data ($x_{\text{obs}}$, $y_{\text{obs}}$) to obtain estimated coefficients $\hat{\beta}$, then used these coefficients to compute predicted values for the validation set. 

By regressing these predictions against true values, we obtain $\lambda_y$ that quantifies the attenuation bias in the label $y$. The results demonstrate how the magnitude of measurement uncertainties affects predictions: with $\sigma_{\text{range}}/\sigma_x = 10$, the fitted regression line shows only 1\% bias ($\lambda_y = 0.99$) compared to the perfect prediction line. When measurement uncertainties become comparable to the signal variance ($\sigma_{\text{range}}/\sigma_x = 1$), we observe substantial attenuation ($\lambda_y = 0.5$) where extreme values are systematically pulled toward the mean. Recall that, for linear regression, the theoretical derivation shows that the attenuation factor for the coefficient $\lambda_\beta$ coincides with the attenuation factor for the predictions $\lambda_y$.

We further validated the theoretical predictions across a range of measurement uncertainties, as shown in Figure~\ref{fig:attenuation_factor}. For the attenuation in the regression coefficient ($\lambda_\beta$), we performed 100 independent simulations at each error level and calculated the mean $\hat{\beta}$ to obtain robust statistics. While error bars from these independent simulations are included in the plot, they are too small to be visible. The figure shows that for the linear case, the simulated attenuation factors (blue points) match the theoretical predictions (blue solid line) exactly across all values of $\sigma_{\text{range}}/\sigma_x$, confirming the analytical derivation.

Through additional simulations varying both $x$ and $y$ measurement uncertainties, we confirmed that the attenuation bias depends solely on the ratio $\sigma_{\text{range}}/\sigma_x$ while remaining independent of measurement uncertainty in $y$, exactly as predicted by our theoretical framework.

The agreement between simulation and theory confirms that for univariate linear regression with known $\sigma_x$, we can predict the attenuation in both coefficients and predictions regardless of the uncertainty in the dependent variable. However, as we will demonstrate below, such calibration against this fundamental bias becomes highly challenging, if not impossible, in higher-order polynomial regression or multivariate cases for linear regression.

\subsection{Attenuation Bias for Higher-Order Regression}
\label{sec:higher-order-regression}

While the theoretical framework precisely describes attenuation bias in linear regression, many astronomical applications, particularly in spectral analysis, employ more complex models \citep{Ting2019,Fabbro2018,Leung2019,OBriain2021}. For example, The Cannon employs quadratic spectral features, while The Payne and AstroNN use neural networks. Empirically, attenuation bias has been observed across various machine learning applications in astronomy, from generalized linear regression with extended features to more sophisticated approaches \citep[see for example fig 7 and 8 of][]{Xiang2019}. Different applications exhibit varying degrees of attenuation bias, raising the question of how our framework extends to these cases.

Many machine learning methods can be viewed as linear regression models with transformed features. Neural networks, for instance, act as feature extractors where intermediate layers perform nonlinear transformations of the input before a final linear transformation layer. Similarly, Gaussian Processes can be interpreted as extended linear regression models where the kernel trick substitutes for explicit feature transformations. While this suggests our theory should extend to these methods, a key challenge emerges: for transformed features, even in simple polynomial regression, the relationship between measurement uncertainties and true scatter becomes non-trivial.

To develop intuition for this complexity, we examine the univariate case with polynomial features, where the relationship between observed quantities and physical parameters often follows higher-order dependencies. Consider the transformation $x\rightarrow x^n$ with the model:
\begin{equation}
y_{\text{true}} = \beta x_{\text{true}}^n,
\end{equation}
\begin{equation}
y_{\text{obs}} = y_{\text{true}} + \delta_y,
\end{equation}
where $\beta$ is the coefficient and $\delta_y$ represents measurement uncertainty in $y$. As before, we assume:
\begin{equation}
x_{\text{obs}} = x_{\text{true}} + \delta_x, \quad \delta_x \sim \mathcal{N}(0, \sigma_x^2).
\end{equation}

While this remains a linear regression in terms of $x^n$, the attenuation bias now depends on the ratio between the range of $x^n$ (denoted $\sigma_{\text{range}, x^n}$) and its measurement uncertainty ($\sigma_{x^n}$). To estimate this ratio, we employ logarithmic variance analysis. Assuming small measurement errors ($\delta_x \ll x_{\text{true}}$), we can approximate:
\begin{equation}
\log x_{\text{obs}} = \log(x_{\text{true}} + \delta_x) \approx \log x_{\text{true}} + \frac{\delta_x}{x_{\text{true}}}.
\end{equation}

\begin{figure}
    \centering
    \includegraphics[width=0.45\textwidth]{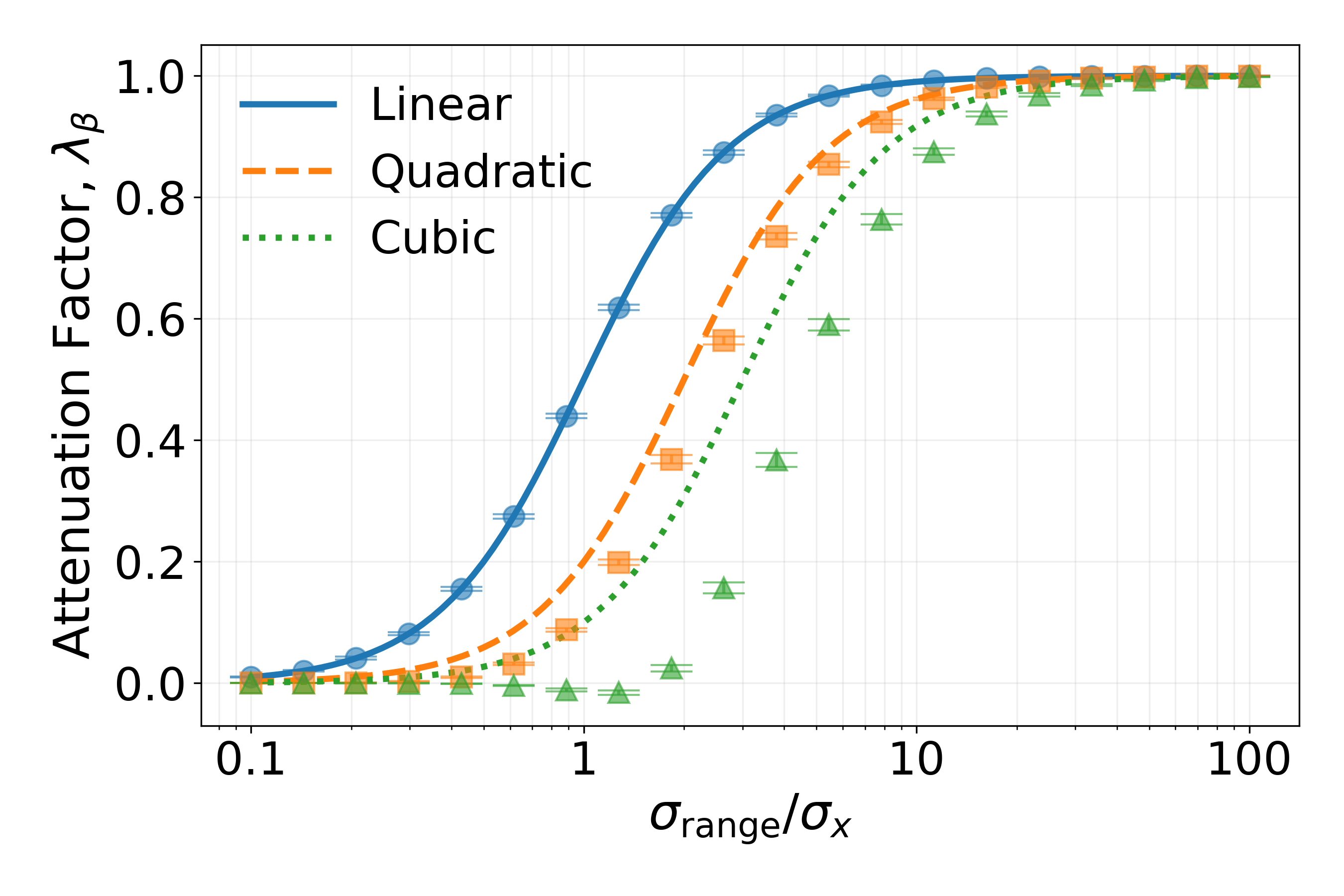}
    \caption{Impact of measurement uncertainties on coefficient attenuation bias in univariate regression. The attenuation factor for the regression coefficient ($\lambda_\beta$) as a function of $\sigma_{\text{range}}/\sigma_x$ shows analytical predictions (lines) and numerical simulations (points) for linear (blue-solid), quadratic (orange-dashed), and cubic (green-dotted) terms. For each polynomial order, $\lambda_\beta$ is defined as the attenuation bias factor of the highest order coefficient.}
    \label{fig:attenuation_factor}
 \end{figure}

The variance propagates as:
\begin{equation}
\operatorname{Var}(\log x_{\text{obs}}) \approx \left( \frac{\sigma_x}{x_{\text{true}}} \right)^2 \approx \left( \frac{\sigma_x}{\sigma_{\text{range}}} \right)^2,
\end{equation}
where we approximate $x_{\text{true}} \approx \sigma_{\text{range}}$. For the $n$-th power:
\begin{equation}
\operatorname{Var}(\log x_{\text{obs}}^n) = n^2 \operatorname{Var}(\log x_{\text{obs}}) = n^2 \left( \frac{\sigma_x}{\sigma_{\text{range}}} \right)^2.
\end{equation}

This leads to an approximation for the ratio of uncertainties in transformed space:
\begin{equation}
\left( \frac{\sigma_{x^n}}{\sigma_{\text{range},x^n}} \right)^2 \simeq n^2 \left( \frac{\sigma_x}{\sigma_{\text{range}}} \right)^2,
\end{equation}
yielding an attenuation factor for the $n$-th degree transformation:
\begin{equation}
\lambda_{\beta,n} = \frac{1}{1 + n^2 \left( \dfrac{\sigma_x}{\sigma_{\text{range}}} \right)^2}.
\end{equation}

This result reveals that higher-order terms suffer more severe attenuation bias: for the same measurement uncertainty $\sigma_x$, the attenuation factor decreases roughly quadratically with the polynomial degree $n$. This has important implications for astronomical applications where higher-order relationships are common.

Figure~\ref{fig:attenuation_factor} illustrates how attenuation bias scales with measurement uncertainty across different functional relationships, comparing the theoretical predictions with numerical simulations. In our simulations, we generated 10,000 samples for each case and centered the $x$ variables to minimize higher-order effects in the Taylor expansion approximation. For each measurement uncertainty level, as before, we performed 100 independent simulations to obtain robust statistics.

As shown in the figure, at any fixed measurement uncertainty, higher-order terms exhibit progressively stronger attenuation bias, evident in the systematic downward shift of the curves with increasing polynomial degree.  Even at modest measurement uncertainties ($\sigma_{\text{range}}/\sigma_x = 10$), the attenuation becomes more substantial: while a linear relationship experiences about 1\% attenuation ($\lambda_{\beta,1} = 0.99$), a quadratic relationship shows approximately 4\% attenuation ($\lambda_{\beta,2} = 0.96$), and a cubic relationship exhibits nearly 8\% attenuation ($\lambda_{\beta,3} = 0.92$).

Also notably, for linear regression ($n=1$), the simulated attenuation factors (blue points) match the theoretical predictions (blue solid line) exactly across all values of $\sigma_{\text{range}}/\sigma_x$. However, for higher-order polynomials ($n=2,3$), while the simulations follow the general trend predicted by theory, they show increasing deviation from theoretical predictions (orange dashed and green dotted lines) as measurement uncertainties become larger ($\sigma_{\text{range}}/\sigma_x < 1$). The simulations reveal that the actual attenuation effects are even stronger than predicted by our simplified theoretical model in these regimes. These deviations are expected because our non-linear approximation above relies on a Taylor expansion that assumes $\sigma_x/\sigma_{\text{range}}$ is small, allowing us to neglect higher-order terms. When this assumption breaks down, additional terms in the expansion become significant, leading to the observed discrepancies.

Since most sophisticated methods involve highly nonlinear transformations of input features, their attenuation bias cannot be analytically predicted or corrected and appears to be a fundamental limitation of data-driven models.

\subsection{Theoretical Framework - Multivariate Linear Regression}

Having established the principles of attenuation bias in the univariate case, we now extend our analysis to multiple independent variables. This extension is particularly relevant for astronomical applications, where observations often involve high-dimensional data such as spectra with numerous wavelength pixels. While linear regression is widely applied in spectral analysis \citep{Ness2015,Ting2016,Rix2016,Hogg2019}, we will show that without proper treatment, it can introduce non-negligible bias in downstream scientific tasks.

Consider a multivariate linear regression model with $p$ variables, where $p$ represents the number of pixels. As before, we absorb the intercept into the feature vector and write:
\begin{equation}
    y_{\text{true}} = \sum_{j=1}^{p} \beta_j x_{\text{true},j}  =   \mathbf{\beta}^\top\mathbf{x}_{\text{true}} 
\end{equation}
\begin{equation}
    y_{\text{obs}} = y_{\text{true}} + \delta_y
\end{equation}
where $\beta_j$ are the true coefficients and $\delta_y$ represents the combined measurement uncertainty and intrinsic scatter in $y$, with $\mathbb{E}[\delta_y] = 0$ and $\text{Var}(\delta_y) = \sigma^2_{y}$. Since uncertainties in $y$ do not interact with the independent variables, they have identical effects on the regression analysis and do not influence the attenuation bias, as demonstrated in our univariate case.

In spectral analysis using discriminative models, each $x_j$ represents a different wavelength pixel in the spectrum. For our analysis, we consider continuum-normalized flux values that contain distinct spectral features, while the dependent variable $y$ represents the physical parameter we aim to infer, such as stellar metallicity [Fe/H]. And as in the univariate case, we assume each flux pixel is observed with measurement uncertainty:
\begin{equation}
    x_{\text{obs},j} = x_{\text{true},j} + \delta_{x,j}, \quad \delta_{x,j} \sim \mathcal{N}(0, \sigma_{x_j}^2)
\end{equation}

For simplicity, we assume that each normalized flux $x_j$ has uncorrelated and unbiased measurement uncertainties with equal variance, i.e., $\delta_{x,j} \sim \mathcal{N}(0,\sigma^2_{x})$ where $\delta_{x,j}$ are independent with $\mathbb{E}[\delta_{x,j}] = 0$ and $\operatorname{Var}(\delta_{x,j}) = \sigma_x^2$. While this is a simplification, it is reasonable for spectroscopic data where different wavelength bins often have similar noise properties due to common instrumental and observational conditions \citep{Nidever2015,Majewski2017,Wilson2019,Guy2023}.

For multivariate linear regression minimizing mean-squared error, recall that the ordinary least squares (OLS) estimator has the analytic form:
\begin{equation}
    \mathbb{E}[\hat{\boldsymbol{\beta}}] = (\mathbf{X}_{\text{obs}}^\top \mathbf{X}_{\text{obs}})^{-1} \mathbf{X}_{\text{obs}}^\top \mathbf{y}_{\text obs}
    \label{eq:ols_estimator}
\end{equation}

To analyze how measurement uncertainties affect this estimator, we must consider the relationship between $\mathbf{X}_{\text{obs}}$ and $\mathbf{X}_{\text{true}}$. Let $\boldsymbol{\Delta}_x$ be the $n \times p$ matrix of measurement errors where each element $\delta_{x,ij}$ represents the measurement error for the $i$-th observation of the $j$-th input, with $n$ being the number of training samples and $p$ the number of features.

For large $n$ or small measurement uncertainties, we can use the first-order approximation:
\begin{equation}
\mathbb{E}[\hat{\boldsymbol{\beta}}] \approx (\mathbb{E}[\mathbf{X}_{\text{obs}}^\top \mathbf{X}_{\text{obs}}])^{-1} \mathbb{E}[\mathbf{X}_{\text{obs}}^\top \mathbf{y}_{\text obs}]
\label{eq:expectation_approximation_scaled}
\end{equation}

First, we compute $\mathbb{E}[\mathbf{X}_{\text{obs}}^\top \mathbf{X}_{\text{obs}}]$ by expanding $\mathbf{X}_{\text{obs}} = \mathbf{X}_{\text{true}} + \boldsymbol{\Delta}_x$:
\begin{align}
\mathbb{E}[\mathbf{X}_{\text{obs}}^\top \mathbf{X}_{\text{obs}}] &= \mathbb{E}[(\mathbf{X}_{\text{true}} + \boldsymbol{\Delta}_x)^\top (\mathbf{X}_{\text{true}} + \boldsymbol{\Delta}_x)] \nonumber \\
&= \mathbb{E}[\mathbf{X}_{\text{true}}^\top \mathbf{X}_{\text{true}}] + \mathbb{E}[\mathbf{X}_{\text{true}}^\top \boldsymbol{\Delta}_x] \nonumber \\
& \qquad + \mathbb{E}[\boldsymbol{\Delta}_x^\top \mathbf{X}_{\text{true}}] + \mathbb{E}[\boldsymbol{\Delta}_x^\top \boldsymbol{\Delta}_x]
\end{align}

Using $\mathbb{E}[\boldsymbol{\Delta}_x] = 0$ and independence of $\boldsymbol{\Delta}_x$ and $\mathbf{X}_{\text{true}}$, the cross terms vanish. For $\mathbb{E}[\boldsymbol{\Delta}_x^\top \boldsymbol{\Delta}_x]$, since measurement errors are uncorrelated across pixels and observations with variance $\sigma_x^2$, this matrix multiplication yields a diagonal matrix where each diagonal element is the sum of $n$ squared errors, giving us $n\sigma_x^2$ for each diagonal element. Hence:
\begin{equation}
\mathbb{E}[\mathbf{X}_{\text{obs}}^\top \mathbf{X}_{\text{obs}}] = \mathbb{E}[\mathbf{X}_{\text{true}}^\top \mathbf{X}_{\text{true}}] + n\sigma_x^2 \mathbf{I}_p
\label{eq:expected_XtX_scaled}
\end{equation}

Next, we compute $\mathbb{E}[\mathbf{X}_{\text{obs}}^\top \mathbf{y}_{\text obs}]$:
\begin{align}
\mathbb{E}[\mathbf{X}_{\text{obs}}^\top \mathbf{y}_{\text obs}] &= \mathbb{E}[(\mathbf{X}_{\text{true}} + \boldsymbol{\Delta}_x)^\top (\mathbf{X}_{\text{true}} \boldsymbol{\beta} + \delta_y)] \nonumber \\
&= \mathbb{E}[\mathbf{X}_{\text{true}}^\top \mathbf{X}_{\text{true}} \boldsymbol{\beta} + \mathbf{X}_{\text{true}}^\top \delta_y \nonumber \\
& \qquad \qquad + \boldsymbol{\Delta}_x^\top \mathbf{X}_{\text{true}} \boldsymbol{\beta} + \boldsymbol{\Delta}_x^\top \delta_y]
\end{align}
Taking expectations and noting that cross-terms involving $\delta_y$ and $\boldsymbol{\Delta}_x$ have zero expectation:
\begin{equation}
\mathbb{E}[\mathbf{X}_{\text{obs}}^\top \mathbf{y}_{\text{obs}}] = \mathbb{E}[\mathbf{X}_{\text{true}}^\top \mathbf{X}_{\text{true}}] \boldsymbol{\beta}
\label{eq:expected_Xty_scaled}
\end{equation}

Therefore best estimate of the multidimensional $\hat{\beta}$ simplifies as:
\begin{equation}
\mathbb{E}[\hat{\boldsymbol{\beta}}] = (\mathbb{E}[\mathbf{X}_{\text{true}}^\top \mathbf{X}_{\text{true}}] + n\sigma_x^2 \mathbf{I}_p)^{-1} \mathbb{E}[\mathbf{X}_{\text{true}}^\top \mathbf{X}_{\text{true}} ]\boldsymbol{\beta}
\label{eq:expectation-beta-multivariate}
\end{equation}

To further simplify the expression, we examine two distinct cases, beginning with independent features.

\subsubsection{Case 1: Independent Multivariate Features}
\label{sec:multivariate-theory-indepedent}

Let's first examine the case where the features $x_{\text{true},j}$ are mutually independent. This scenario commonly occurs when different factors contribute additively to an outcome without interacting with each other. When features are independent, $\mathbb{E}[\mathbf{X}_{\text{true}}^\top \mathbf{X}_{\text{true}}]$ simplifies to a diagonal matrix where each diagonal element represents the variance of the corresponding feature multiplied by the sample size:
\begin{equation}
    \mathbb{E}[\mathbf{X}_{\text{true}}^\top \mathbf{X}_{\text{true}}] = n\, \text{diag}(\sigma^2_{\text{range},1}, ..., \sigma^2_{\text{range},p})
\end{equation}
where $\sigma^2_{\text{range},j}$ represents the variance of the true signal in the $j$-th feature. Substituting this into our previous derivation:
\begin{equation}
    \mathbb{E}[\hat{\boldsymbol{\beta}}] = (n \, \text{diag}(\sigma^2_{\text{range},j}) + n \sigma_x^2 \mathbf{I}_p)^{-1} (n\,\text{diag}(\sigma^2_{\text{range},j})) \boldsymbol{\beta}
    \label{eq:expected_beta}
\end{equation}

This diagonal structure, combined with the independence of measurement errors across pixels, leads to a simple form for each coefficient's expectation:
\begin{equation}
    \mathbb{E}[\hat{\beta}_j] = \frac{\sigma^2_{\text{range},j}}{\sigma^2_{\text{range},j} + \sigma^2_x} \beta_j = \lambda_{\beta,j} \beta_j
\end{equation}

This derivation reveals that, for independent features, the multivariate case preserves the fundamental characteristics of attenuation bias observed in our univariate analysis. Each coefficient experiences attenuation governed by the ratio of its feature's true variance to measurement uncertainty, suggesting that the underlying mechanism of bias remains unchanged in higher dimensions, though the magnitude may vary across features depending on their individual signal variance.

Further, for the variance of the multivariate coefficient estimator, we can extend our univariate analysis using similar reasoning. For independent features and if both $y$ and ${\bf x}$ are centered (mean substracted), and ${\bf x}$ have the same signal range ${\sigma_{\rm range}}$ and measurement uncertainty $\sigma_x$,  (see the derivation in Appendix~\ref{app:variance}), the variance of each coefficient estimate follows the same form as the univariate case:
\begin{equation}
  \text{Var}(\hat{\beta}_j) = \frac{\sigma^2_y + \sum_{k=1}^p \beta_k^2\sigma^2_{x} }{n(\sigma^2_{\text{range}} + \sigma^2_{x})} \label{eq:variance-multivariate}
\end{equation}

However, as established in our univariate analysis, in many astronomical applications it is more critical to understand the attenuation in terms of the predicted label $y$ rather than the bias in individual coefficients. This is particularly relevant when these coefficients serve as the ``machine" in a data-driven model.

For the multivariate case, the relationship between coefficient attenuation and prediction bias follows directly from our analysis. For any new observation $\mathbf{x}_{\text{obs}}$, the predicted value is:
\begin{equation}
  y_{\text{pred}} = \hat{\boldsymbol{\beta}}^\top\mathbf{x}_{\text{obs}} 
\end{equation}
The expectation of $y_{\text{pred}}$ can be derived by considering that $\hat{\boldsymbol{\beta}}$ is determined from the training set while $\mathbf{x}_{\text{obs}}$ comes from new observations, making them statistically independent. Additionally, since measurement errors are assumed to be unbiased, $\mathbb{E}[\mathbf{x}_{\text{obs}}] = \mathbf{x}_{\text{true}}$. Thus:
\begin{align}
  \mathbb{E}[y_{\text{pred}}] &= \mathbb{E}[\hat{\boldsymbol{\beta}}^\top\mathbf{x}_{\text{obs}}] \nonumber \\
  &= \mathbb{E}[\hat{\boldsymbol{\beta}}]^\top\mathbb{E}[\mathbf{x}_{\text{obs}}] \nonumber \\
  &= \mathbb{E}[\hat{\boldsymbol{\beta}}]^\top\mathbf{x}_{\text{true}}
\end{align}

A key limiting case arises when all features have equal signal variance, i.e., $\sigma_{\text{range},j} = \sigma_{\text{range}}$ for all $j$. In this case, the expectation of the predicted value becomes:
\begin{align}
   \mathbb{E}[y_{\text{pred}}] &= \mathbb{E}[\hat{\boldsymbol{\beta}}^\top]\mathbf{x}_{\text{true}} \nonumber \\
   &= \lambda_{\beta} \boldsymbol{\beta}^\top\mathbf{x}_{\text{true}} \nonumber \\
   &= \lambda_{\beta} \sum_{j=1}^p \beta_j x_{\text{true},j} \nonumber \\
   &= \lambda_{\beta} y_{\text{true}}
\end{align}
where we have used that all coefficients share the same attenuation factor $\lambda_{\beta}$ when $\sigma_{\text{range},j} = \sigma_{\text{range}}$ for all $j$, and the last equality follows from the definition of $y_{\text{true}}$. Thus, we recover the same form of attenuation bias as in the univariate case, where the bias in coefficients directly translates to bias in predicted labels.

This result reveals that, for independent features of equal strength, the attenuation bias is independent of the number of features $p$. The bias induced by measurement uncertainties in the input features cannot be mitigated simply by including more features in the model.

For the more general case where $\sigma_{\text{range},j}$ varies across features, the attenuation in $y_{\text{pred}}$ takes a more complex form:
\begin{align}
   \mathbb{E}[y_{\text{pred}}] &= \mathbb{E}[\hat{\boldsymbol{\beta}}^\top]\mathbf{x}_{\text{true}} \nonumber \\
   &= \sum_{j=1}^p \lambda_{\beta,j} \beta_j x_{\text{true},j} \nonumber \\
   &= \sum_{j=1}^p \frac{\sigma^2_{\text{range},j}}{\sigma^2_{\text{range},j} + \sigma^2_x} \beta_j x_{\text{true},j}
\end{align}

Comparing this to $y_{\text{true}} = \sum_{j=1}^p \beta_j x_{\text{true},j}$, we can define an effective attenuation factor for the predicted label:
\begin{equation}
   \lambda_y = \frac{\sum_{j=1}^p \lambda_{\beta,j} \beta_j x_{\text{true},j}}{\sum_{j=1}^p \beta_j x_{\text{true},j}}
\end{equation}

This effective attenuation $\lambda_y$ represents a weighted average of the individual coefficient attenuation factors $\lambda_{\beta,j}$, where the weights depend on the contribution of each feature to $y_{\text{true}}$. For features that carry no information (e.g., continuum pixels with $\sigma_{\text{range},j} \approx 0$), both $\lambda_{\beta,j}$ and $\beta_j$ approach zero, effectively removing their contribution to the overall bias. For the informative features where signal are comparable, i.e.  $\beta_j x_{\text{true},j}$ is about the same for all $j$, then $\lambda_y$ approaches the arithmetic mean of $\lambda_{\beta,j}$ over the informative pixels (see Appendix~\ref{app:independent-mean}).

\begin{figure*}[htbp]
    \centering
    \includegraphics[width=\textwidth]{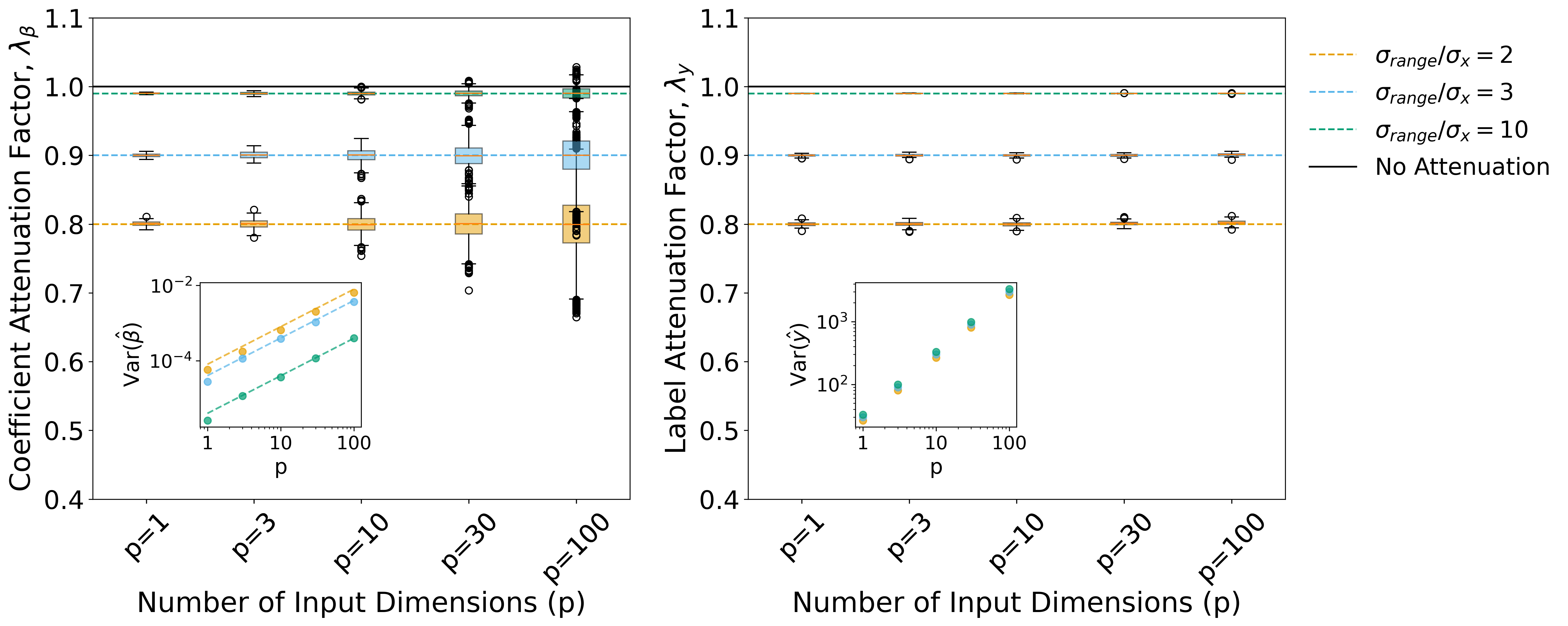}
    \caption{Effect of dimensionality on attenuation bias in coefficients and predictions for independently drawn input features in multivariate linear regression. Left panel shows the distribution of coefficient attenuation factors ($\lambda_{\beta} = \hat{\beta}/\beta_{\text{true}}$), pooling results across all coefficients and simulations for each $p$, with boxcars showing the distribution over all coefficients and 100 simulations. Right panel shows prediction attenuation factors ($\lambda_y$), obtained by regressing predicted values against true values across all samples in each simulation, with boxcars showing the distribution over all simulations. For both panels, we examined three different signal-variance-to-noise ratios ($\sigma_{\text{range}}/\sigma_x = 2, 3, 10$). The boxes indicate the interquartile range (IQR) with the median shown as a horizontal line, and whiskers extend to points within 1.5×IQR. The open circles show individual realizations that fall beyond the whiskers. The dashed lines show the theoretical predictions of $\lambda = 1/(1 + (\sigma_x/\sigma_{\text{range}})^2)$ for each signal-variance-to-noise ratio. The inset plots show the variance of the coefficient estimates (left) and predicted values (right) as a function of dimensionality, with points showing empirical results and dashed lines showing theoretical predictions.}
    \label{fig:dimension_effect}
\end{figure*}

\subsubsection{Case 2: Correlated Multivariate Features}
\label{sec:multivariate-correlated-theory}

In most astronomical applications, there is still a saving grace stemming from the fact that it is the label $y$, the physical property of the object, that generates all the features $\mathbf{x}$, i.e., $y$ causes $\mathbf{x}$. By definition, in many astronomical applications, the features $\mathbf{x}$ are not independent (their measurement errors can be independent, but not $\mathbf{x}_{\text{true}}$ themselves). As we will demonstrate below, such correlation in features helps mitigate the attenuation bias as the number of features increases. 

This is perhaps not surprising; intuitively, with multiple correlated features, the noise needs to conspire in a single direction to generate the bias. As the dimensionality increases, the likelihood of such action happening becomes smaller for any independent random noise.

For an arbitrary set of features with non-zero correlation, an analytic theory for the attenuation bias becomes intractable. This is another reason why such attenuation bias, while fundamental to many data-driven machine learning applications as we have demonstrated, is hard to eradicate from first principles in real practice.

To derive some analytic expressions and gain insights for this more general case, we will make the simplifying assumption of an optimal scenario where all features are completely correlated with a correlation of one. In this case, all features can be expressed as linear transformations of a single underlying variable $x_{\text{true}}$,  we can thus assume $\mathbf{x}_{\text{true},j} = a_j x_{\text{true}}$ are the true independent variables, with $a_j$ being known scaling factors. The scaling factors $a_j$ capture the relative magnitudes of each feature with respect to this common variable. For example, in spectral models, $a_j$ approximately corresponds to the oscillator strength when all features lie within the linear portion of the curve of growth.

If we further assume that $\boldsymbol{\beta} = \beta \mathbf{a}$, where $\beta$ is a scalar coefficient, i.e., the linear coefficients follow also the direction of the scaling factors, as we will show in Appendix~\ref{app:multivariate}, we can still analytically derive the attentuation bias factor for the regression coefficients. The expected value of the coefficient estimates $\hat{\boldsymbol{\beta}}$ is approximately:
\begin{equation}
\mathbb{E}[ \hat{\boldsymbol{\beta}} ] \approx \lambda_\beta \mathbf{\beta},
\label{eq:expected_beta_final_scaled}
\end{equation}
where the attenuation factor $\lambda_\beta$ is:
\begin{equation}
\lambda_\beta = \frac{ \sum_{j=1}^{p} a_j^2 }{ \frac{ n \sigma_x^2 }{ S_{x} } + \sum_{j=1}^{p} a_j^2 }.
\end{equation}
Here, $\sigma_x^2$ is the measurement error variance, $S_{x} = \sum_{i=1}^n x_{\text{true},i}^2$ is the sum of squares of the true signal, $n$ is the number of samples, and $p$ is the number of features.

As shown in the equation, similar to the univariate case, the attenuation bias in the correlated case depends on the ratio of the total measurement error variance ($n \sigma_x^2$) to the total signal variance indicator ($S_{x}$). However, it also depends on the scaling factors $a_j$ and, more importantly, the number of variables $p$, which was not the case before. If we define an effective signal range vs measurement error as:
\begin{equation}
\sigma' = \sqrt{\frac{S_{x}}{n \sigma_x^2} \sum_{j=1}^{p} a_j^2},
\label{equation:attenuation-factor-multivariate-1}
\end{equation}
the attenuation factor can be rewritten in a more familiar form as before:
\begin{equation}
\lambda_\beta = \frac{1}{1 + \frac{1}{\sigma'^2}}.
\label{equation:attenuation-factor-multivariate-2}
\end{equation}

Interestingly, unlike the independent feature case, for a large number of correlated and non-trivial features ($p \to \infty$), we have $\sigma' \to \infty$, and the attenuation factor $\lambda_\beta$ approaches 1, indicating that the bias can be mitigated by increasing the dimensionality of the input space, provided the features are sufficiently correlated. The attenuation bias in the correlated case is less severe than in the independent case, as the correlation among features allows the model to leverage additional information to reduce the impact of measurement errors.

\subsection{Numerical Demonstrations -- Multivariate Linear Regression}

\subsubsection{Case 1: Independent Multivariate Features}

To illustrate the theoretical framework of multivariate attenuation bias, we conducted numerical simulations examining both coefficient and prediction attenuation across varying numbers of input dimensions. We will first focus on the case where the features independently contribute to the output label.

We generated synthetic data following our established framework in the case of univariate: for each j-th dimension (``pixel") of the input, we sample true values uniformly from [-5, 5] (yielding $\sigma_{\text{range}} \approx 2.9$) and adding homogeneous Gaussian measurement errors with a variance of $\sigma_x^2$ independent of $j$. We first draw the values of the input ${\bf x}$. The value of all pixels are drawn independently, and their linear sum from the noiseless version with the predefined coefficient is then assumed to be the true labels. 

To explore the relationship between measurement uncertainty and attenuation bias, we examined three different signal variance to measurement error ratio for the input ($\sigma_{\text{range}}/\sigma_x = 2, 3, 10$). For each dimension $p$ ranging from 1 to 100, we performed 100 independent simulations with 10,000 samples each to ensure robust statistics. In each simulation, we maintained identical statistical properties across all dimensions (uniform distribution, consistent measurement error in $\mathbf{x}$ and zero measurement error in $y$) and set all coefficients $\beta_j$ to 2.0, ensuring equal contribution to the dependent variable.

Figure \ref{fig:dimension_effect} presents our findings through a two-panel visualization comparing coefficient and prediction attenuation factors across different numbers of input dimensions. For each simulation, we first fit a linear regression model using the noised-up observed $\mathbf{X}_{\text{obs}}$ to predict the output label $y$, obtaining estimates $\hat{\boldsymbol{\beta}}$ for the true coefficients $\boldsymbol{\beta}$. To quantify the coefficient attenuation, we calculate the ratio $\lambda_{\beta} = \hat{\beta}/\beta_{\text{true}}$ for each coefficient. For prediction attenuation, we use the fitted model to make predictions on the observed features and then perform a linear regression of these predictions against the true values, where the slope of this regression gives us $\lambda_y$. The left panel shows the distribution of coefficient attenuation factors ($\lambda_{\beta}$), pooling results across all coefficients and simulations for each $p$. The boxcar shows the distribution of $\lambda_\beta$ over all $p$ coefficients and all simulations. The right panel displays prediction attenuation factors ($\lambda_y$), obtained by regressing predicted values against true values across all samples in each simulation, and the boxcar shows the distribution of $\lambda_y$ over all simulations.

\begin{figure}[htbp]
    \centering
    \includegraphics[width=0.48\textwidth]{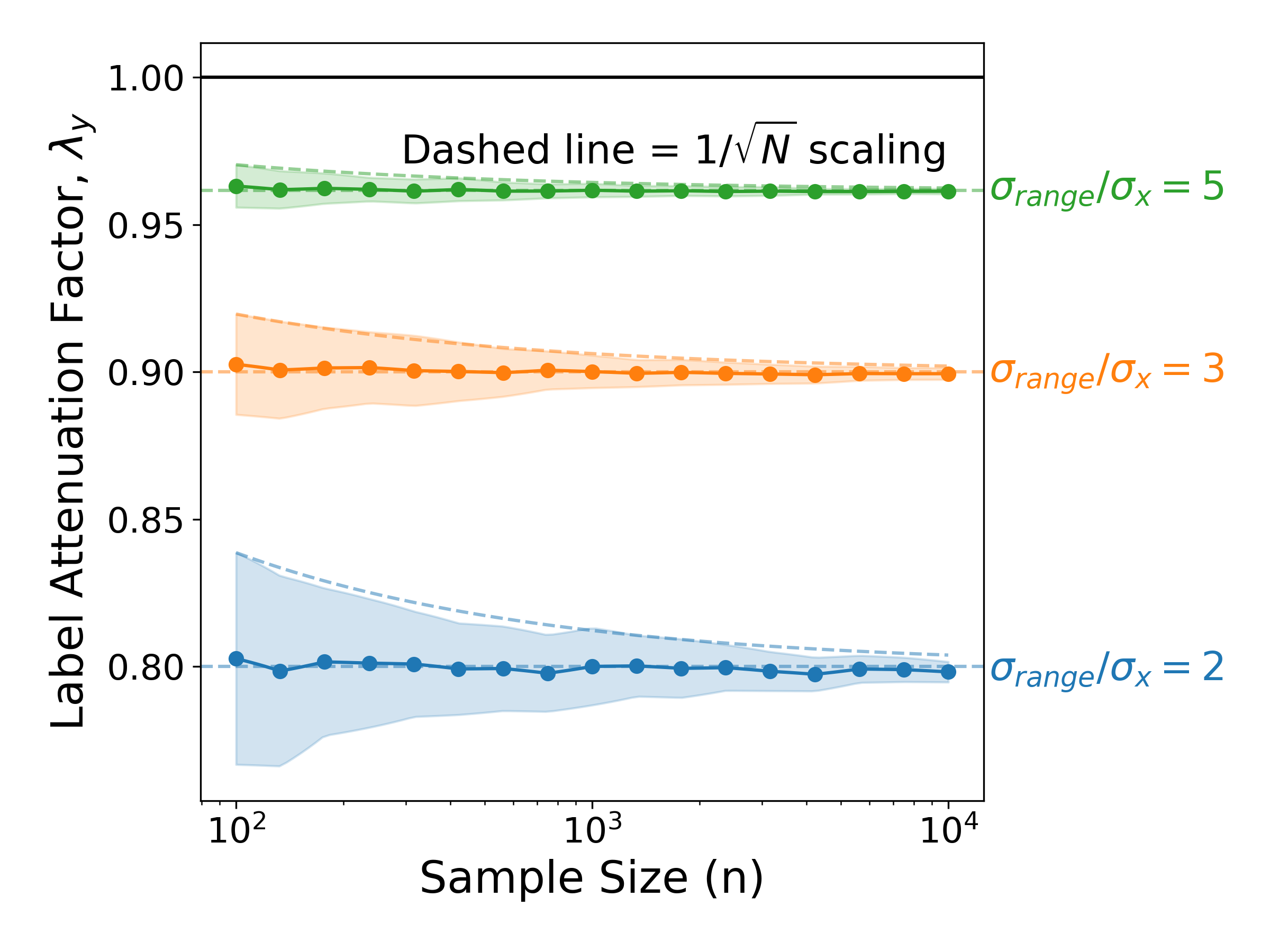}
    \caption{Prediction attenuations versus training sample size for different ratios of signal range to measurement error with fixed input dimension $p=5$ for multivariate linear regression with independent input features. Solid lines show mean attenuation factors ($\lambda_y$) with shaded regions indicating $\pm 1$ standard deviation calculated from 100 independent simulations. Results shown for three ratios ($\sigma_{\text{range}}/\sigma_x = 2, 3, 5$). For each ratio, horizontal dashed line indicates theoretical attenuation limit $\lambda = 1/(1 + (\sigma_x/\sigma_{\text{range}})^2)$, curved dashed lines show $1/\sqrt{n}$ convergence scaling from coefficient estimate variance. Solid black line at $\lambda_y = 1$ indicates case of no attenuation.}
    \label{fig:sample_dimension_effect}
\end{figure}

\begin{figure*}[htbp]
    \centering
    \includegraphics[width=\textwidth]{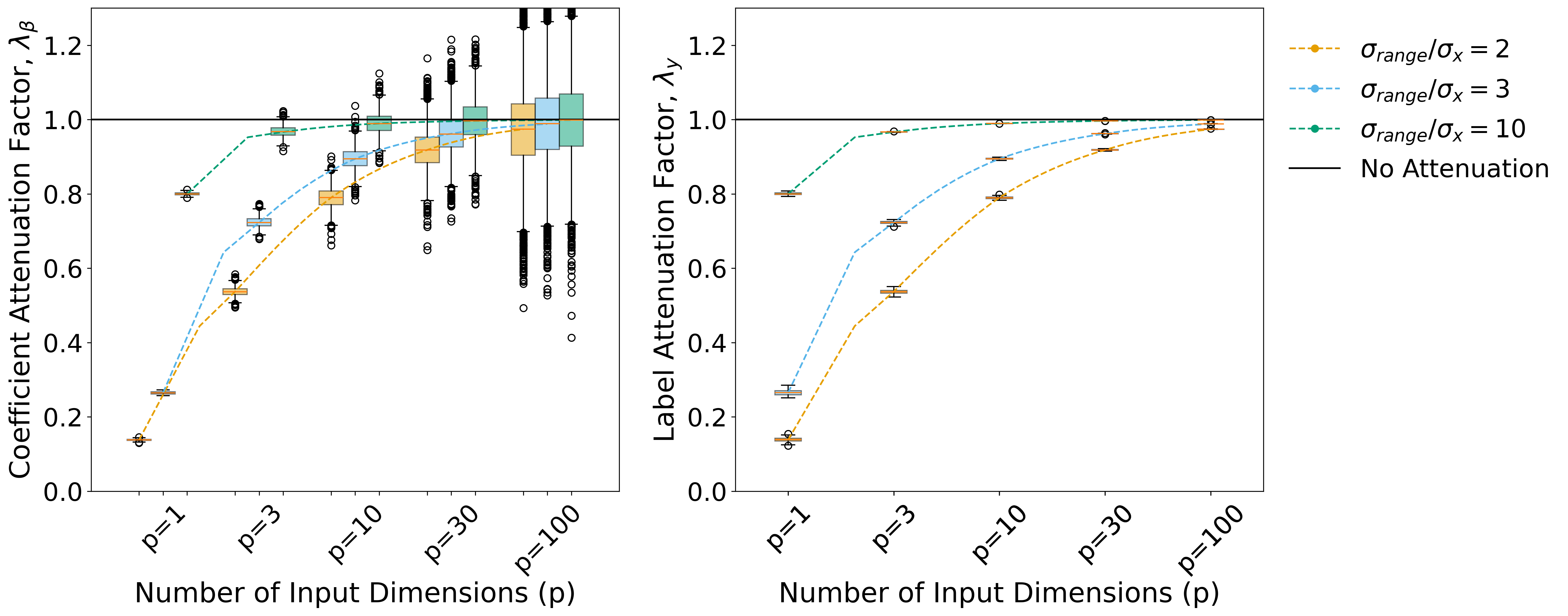}
    \caption{Effect of dimensionality on attenuation bias for perfectly correlated features. Left panel shows coefficient attenuation factors ($\lambda_{\beta}$) and right panel shows prediction attenuation factors ($\lambda_y$) for different numbers of perfectly correlated input dimensions. Results are shown for three ratios of signal range to measurement error ($\sigma_{\text{range}}/\sigma_x = 2, 3, 10$) based on 100 independent simulations. Features are generated as linear transformations of a uniform random variable on [0,1] ($\sigma_{\text{range}} = 1/\sqrt{12}$) with scaling factors $a_j$ varying linearly from 0.2 to 0.4 across dimensions, chosen to ensure labels remain of order unity. In the left panel, box plots show the distribution of coefficient attenuation across all coefficients and simulations, with boxes offset horizontally for visibility across different $\sigma_{\text{range}}/\sigma_x$ ratios. The right panel shows the mean prediction attenuation with $\pm 1$ standard deviation across simulations. Dashed lines in both panels show theoretical predictions. The solid black line at $\lambda = 1$ indicates no attenuation.}
    \label{fig:dimension_effect_correlated}
\end{figure*}

Our results, as shown in Figure~\ref{fig:dimension_effect}, demonstrate excellent agreement with these theoretical predictions across all tested input dimensions. The attenuation factors remain stable as the number of input dimensions increases, confirming our analytical finding that for independent features, additional dimensions do not mitigate the attenuation bias. For all three $\sigma_{\text{range}}/\sigma_x$ ratios tested, the observed attenuation closely matches the theoretical prediction of $\lambda = 1/(1 + (\sigma_x/\sigma_{\text{range}})^2)$—the same relationship we derived in the univariate case—shown as dashed lines in both panels. Even in the optimistic case where the signal range is 10 times the measurement error, which represents an optimistic limit for spectral analysis at high-resolution and high-signal-to-noise (see Section~\ref{sec:spectra}), there remains a persistent 1\% systematic bias when the features are independent. Further, the identical behavior between coefficient attenuation ($\lambda_{\beta}$) and prediction attenuation ($\lambda_y$) across all dimensions and signal-to-noise ratios confirms the theoretical framework.

In the case of independent features, not only does increasing the number of pixels fail to mitigate the attenuation bias, it actually comes with a cost. As we derived in Equation~\ref{eq:variance-multivariate}, the variance of the coefficients increases with $p$ due to the $\sum_{k=1}^p \beta_k^2$ term in the numerator. This behavior is perhaps not surprising from a geometric perspective: as dimensionality increases while the number of data points remains constant, we encounter the curse of dimensionality, where estimation of the hyperplane becomes increasingly stochastic. The inset plots in both panels reveal the variance behavior of our estimates as a function of dimensionality. As shown in the inset plot of the left panel, the empirical variance of coefficient estimates (points) closely follows the theoretical prediction (dashed lines) from Equation~\ref{eq:variance-multivariate}, demonstrating how the uncertainty scales systematically with the number of dimensions.

A common suggestion for addressing measurement error effects is to increase the sample size. Having established that increasing the number of input dimensions cannot mitigate the attenuation bias, we now examine how prediction attenuation ($\lambda_y$) depends on training sample size. Following exactly the same simulation setup as in Figure \ref{fig:dimension_effect}, where we generate independent input features drawn from uniform distribution and add measurement errors, we now fix the number of input dimensions ($p=5$) and vary the sample size. As shown in Equation~\ref{eq:variance-multivariate}, the variance of the coefficient estimates scales as $1/n$, where $n$ is the sample size. This fundamental statistical behavior suggests that while the bias remains fixed, the precision of our estimates should improve with larger samples. Figure \ref{fig:sample_dimension_effect} shows this relationship across three different ratios of signal range to measurement error ($\sigma_{\text{range}}/\sigma_x = 2, 3, 5$). 

The results reveal that as sample size increases, the attenuation factor for each $\sigma_{\text{range}}/\sigma_x$ ratio converges to its respective theoretical value at a rate that follows a $1/\sqrt{n}$ scaling law (shown as dashed lines). This scaling behavior directly emerges from the variance of $\hat{\boldsymbol{\beta}}$ in our theoretical derivation and propagates to the predictions through the linear relationship. However, the key finding is that at all training sizes, we observe persistent attenuation in predictions that depends on the ratio $\sigma_{\text{range}}/\sigma_x$. The systematic bias introduced by measurement uncertainties cannot be overcome simply by increasing the number of input dimensions or size of the training sample.

\vspace{1cm}

\subsubsection{Case 2: Correlated Multivariate Features}

While our analysis of independent features presents a concerning picture for attenuation bias, there is reason for cautious optimism. In most astronomical scenarios, such as spectroscopic analysis, the observed features (flux at different wavelengths), treated as the input of discriminative models, are not independent but rather correlated through the underlying physics – it is the physical parameters that generate the spectra, not vice versa. As we have derived in Section~\ref{sec:multivariate-correlated-theory}, this correlation between features could potentially help mitigate attenuation bias.

To explore this scenario analytically, we consider the optimistic limiting case of perfectly correlated features. We generate synthetic data where all features are linear transformations of a single underlying variable $x_{\text{true}}$, drawn from a uniform distribution over [0,1] (yielding $\sigma_{\text{range}} = 1/\sqrt{12}$). Each feature $j$ is scaled by a factor $a_j$, where we choose $a_j$ to vary linearly from 0.2 to 0.4 across the $p$ dimensions. We chose this range of $a_j$, as it ensures the resulting labels $y$ remain roughly of order unity across different $p$, mimicking typical normalized label in spectroscopy analysis as we will describe below (e.g., metallicity).

We have established that deriving the analytic attenuation factor is challenging, if not impossible, for general cases for correlated features. To make the theoretical analysis tractable, as assumed in the theoretical framework, we set the regression coefficients $\boldsymbol{\beta}$ equal to the scaling factors $\mathbf{a}$, allowing us to derive analytical predictions for the attenuation factor of coefficients as shown in Equation~\ref{equation:attenuation-factor-multivariate-1} and \ref{equation:attenuation-factor-multivariate-2}. If we further assume homogeneous and identical (but not independent) $x_{\text{true}}$, the attenuation factor $\lambda_{\beta}$ then translates into the attenuation factor of the label as we take the expectation through the linear transformation. With this framework, we can check with simulations to build insight but to ensure also that the theoretical framework is accurate.

\begin{figure*}[htbp]
    \centering
    \includegraphics[width=\textwidth]{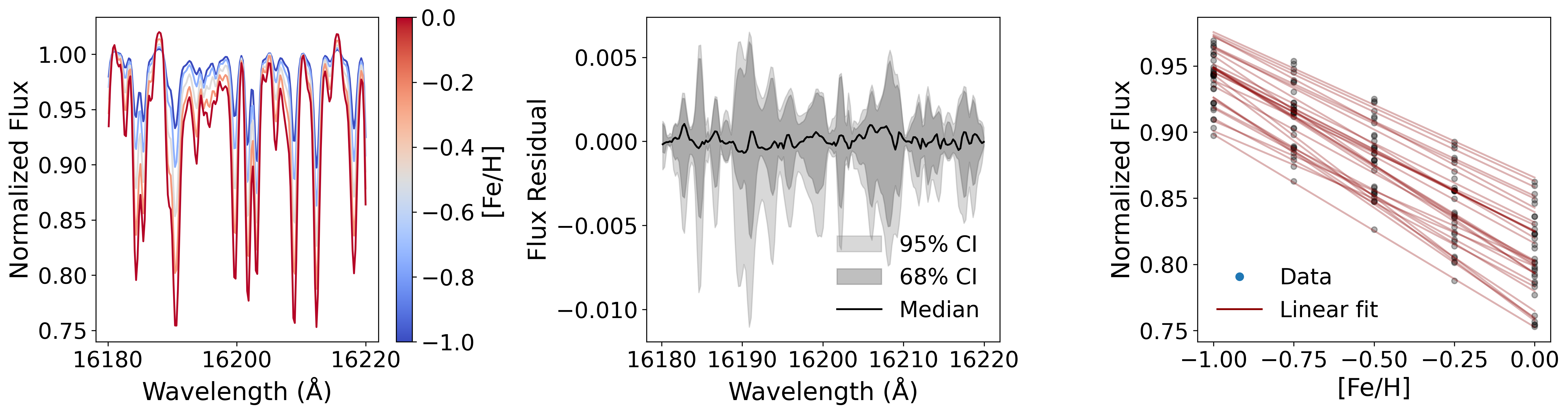}
    \caption{Spectral variations around the Fe absorption line at 16200\AA\ for a red clump star ($T_{\text{eff}} = 4750$K, $\log g = 2.5$) across different metallicities from $\text{[Fe/H]} = -1.0$ to $0.0$. The color gradient indicates metallicity values. Throughout this study, we adopt this 40\AA\ region containing 179 wavelength pixels as our inference regime. Middle panel: The gray bands show $68\%$ (dark) and $95\%$ (light) confidence intervals of flux residuals from linear fits, calculated using 1000 randomly drawn $\text{[Fe/H]}$ values. Right panel: Flux-metallicity relationships for wavelength pixels showing more substantial variation ($>0.1$ in normalized flux comparing the case with $\text{[Fe/H]} = -1$ and $0$) across the $\text{[Fe/H]}$ range. Each set of points represents a different pixel's relationship with $\text{[Fe/H]}$, with red lines showing their linear fits.}
    \label{fig:linear_relation}
\end{figure*}

Figure \ref{fig:dimension_effect_correlated} demonstrates how attenuation bias behaves in this idealized correlated case. As in Figure \ref{fig:dimension_effect}, we perform 100 independent simulations for each dimensionality, but now with perfectly correlated features. In both panels, the markers show the mean attenuation factors, while dashed lines indicate the theoretical predictions. The left panel displays the distribution of coefficient attenuation factors ($\lambda_{\beta}$) through box plots, with boxes showing the interquartile range. The variance is shown for all collected coefficients over all simulations. For better visibility, the box plots for different $\sigma_{\text{range}}/\sigma_x$ ratios are horizontally offset. The right panel shows the prediction attenuation factors ($\lambda_y$) across simulations. The empirical results (box plots) closely match the theoretical predictions (dashed lines) across different $\sigma_{\text{range}}/\sigma_x$ ratios. Like in the case of independent features, the left panel reveals additional complexity through the spread in coefficient attenuation factors $\lambda_{\beta}$, particularly at higher dimensions, suggesting that individual coefficients still suffer from high degrees of variations even when their collective effect is better behaved.

The key difference from the independent case is immediately apparent: increasing the number of correlated features does help mitigate the bias. For $\sigma_{\text{range}}/\sigma_x = 10$, the attenuation factor improves from about 0.8 at $p=1$ to nearly 1.0 at $p=100$. Even for more challenging cases with $\sigma_{\text{range}}/\sigma_x = 2$, we see substantial improvement from 0.13 (almost random prediction) at $p = 1$ to 0.97 at $p = 100$ as dimensionality increases. 

It's worth noting that this represents a best-case scenario for several reasons. First, perfect correlation between features is unrealistic in practice. Second, our assumption that $\boldsymbol{\beta} = \mathbf{a}$ likely overestimates the mitigation effect. In real applications, as the number of features increases, the individual coefficients typically become smaller to maintain reasonable output scales. This means the sum $\sum_{k=1}^p \alpha_k^2$ grows more slowly than in our idealized case, resulting in less mitigation of the attenuation bias as we will see in the real scenario below.

\section{Attenuation Bias in Spectral Analysis}
\label{sec:spectra}

Having established the theoretical framework for attenuation bias in both univariate and multivariate cases, we now examine its implications in a specific astronomical context: the inference of stellar properties from observed stellar spectra. We use APOGEE-like mock spectra as a case study to demonstrate how attenuation bias manifests in spectral inference tasks when training data contains measurement noise.

Spectra present a unique challenge in astronomical data analysis due to their high dimensionality. Each spectrum typically contains thousands of flux measurements across different wavelength pixels, encoding stellar properties through absorption features. Understanding how measurement errors propagate through discriminative machine learning models—which are becoming increasingly prevalent in modern data-driven research—is crucial.

We chose APOGEE as our case study for several reasons. First, we can take advantage of a well-tested spectral emulator which allows us to simulate spectra efficiently. Second, and perhaps more importantly, APOGEE high-resolution stellar spectra represent a conservative test case. With its high resolution ($R = \Delta\lambda/\lambda \simeq 24,000$) and typical signal-to-noise ratio of $\mathcal{O}(100)$, APOGEE should present the most favorable conditions where attenuation bias would be minimal. Stellar features generally exhibit sharper features with more observable variation compared to galaxy spectra or integrated light spectra from stellar populations \citep{Cid-Fernandes1998,Cardiel2003,Dressler2004,Cid-Fernandes2005,Coelho2020}.

However, even under these optimal conditions of APOGEE, attenuation bias remains significant. When varying $\text{[Fe/H]}$ from $-1$ to $0$, only the strongest absorption features show variations of order 10\% in normalized flux, setting a natural scale of $\sigma_{\text{range}} \approx 0.1$. This yields $\sigma_{\text{range}}/\sigma_x \approx 10$. As demonstrated in Section~\ref{sec:theory-univariate}, even this favorable ratio produces an appreciable attenuation bias of 1\% for a single feature. Given that most spectral features show variations significantly smaller than 10\% in flux, attenuation bias remains non-negligible under most assumptions.

Such bias has been broadly observed in many machine learning tasks applied to APOGEE spectra (see for example fig 7-8 in \citealt{Fabbro2018} or fig 9-11 in \citealt{Leung2019}). While much of the consensus often attributes these effects to the precision of training labels or sample size limitations, the theoretical framework and numerical validation show that these factors have no impact on such attenuation bias.

Nonetheless, predicting the attenuation bias in APOGEE spectra requires consideration beyond the simple univariate case, as spectral features in physical systems are inherently correlated. When varying only one label (here $\text{[Fe/H]}$) while fixing other parameters, the information content across pixels should be perfectly correlated through the underlying physics. As developed in Section~\ref{sec:multivariate-correlated-theory}, such correlation between input dimensions (while measurement noise remains independent) should help mitigate the attenuation bias.

The theoretical framework predicts that as the number of correlated features $p$ increases, the attenuation factor should improve, potentially approaching unity as $p \to \infty$. However, the exact analytical prediction relies on several strong assumptions that may not hold in practice. Given these limitations, we must turn to numerical simulations to understand attenuation bias in realistic spectroscopic applications.

\subsection{Numerical Investigation of Attenuation Bias Using Synthetic APOGEE Spectra}
\label{sec:spectra-apogee}

For our analysis, we use The Payne spectral emulation \citep{Ting2019} to generate normalized synthetic APOGEE spectra, though the exact accuracy of the model is not critical to our conclusions about attenuation bias. We chose The Payne emulator because it is publicly available and allows us to efficiently simulate spectra across the APOGEE parameter range. Following APOGEE's wavelength sampling rate, we assume three pixels per resolution element, and all SNR values discussed below refer to the SNR per such pixel.

\begin{figure*}[htbp]
    \centering
    \includegraphics[width=\textwidth]{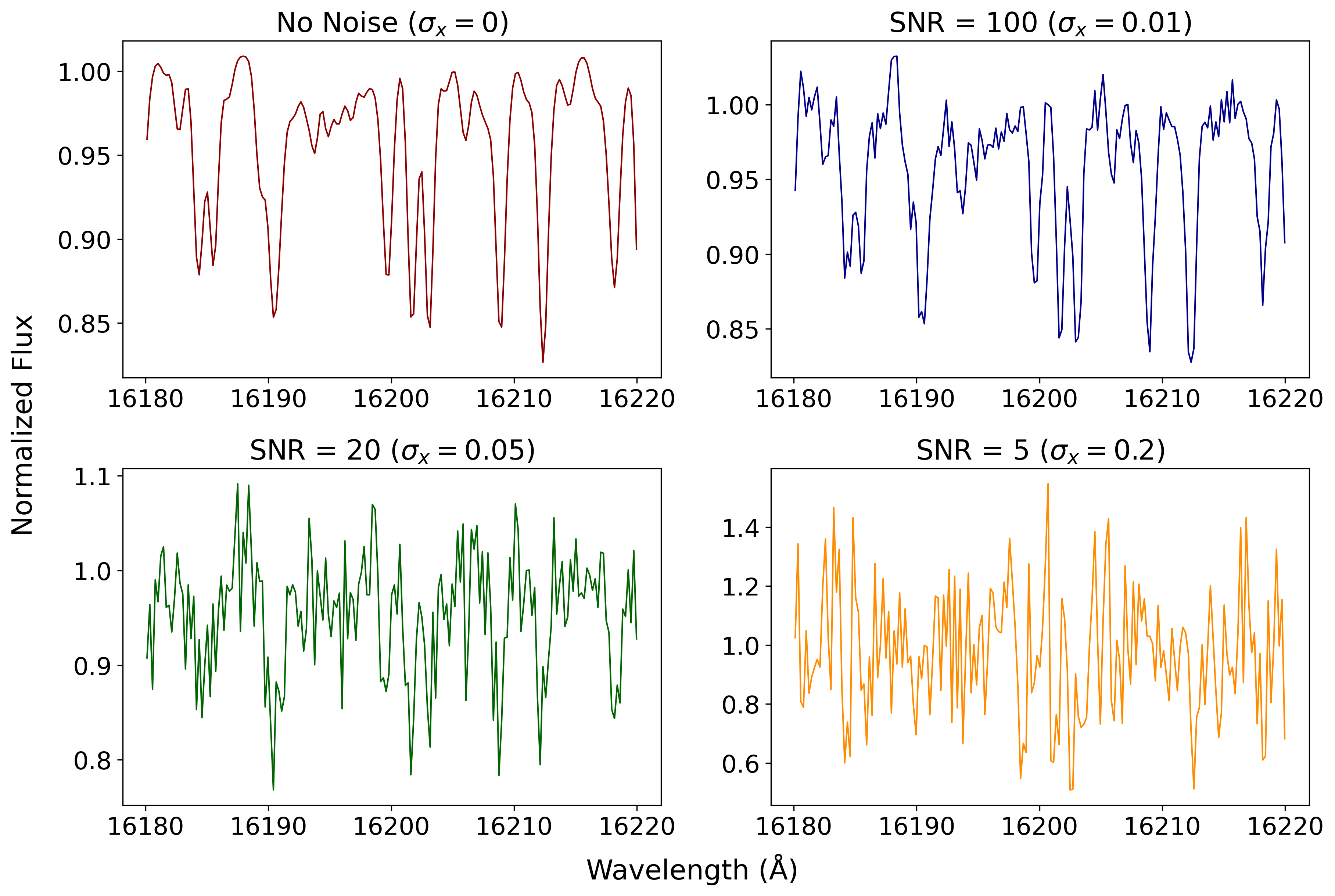}
    \caption{Demonstration of measurement uncertainties in APOGEE-like spectra at 16200\AA\ analyzed in Figure \ref{fig:linear_relation}. All panels show the identical underlying spectrum ($T_{\text{eff}} = 4750$K, $\log g = 2.5$, $\text{[Fe/H]} = -0.5$) but with different measurement uncertainties: no noise ($\sigma_x = 0$, top left), high signal-to-noise ($\sigma_x = 0.01$, SNR = 100, top right), moderate signal-to-noise ($\sigma_x = 0.05$, SNR = 20, bottom left), and low signal-to-noise ($\sigma_x = 0.2$, SNR = 5, bottom right). The range of $\sigma_x$ values chosen spans typical uncertainties in normalized flux from high-resolution surveys like APOGEE, with SNR\,$ \sim 100$, $\sigma_x \sim 0.01$ \citep{Majewski2017,Holtzman2018}, to low-resolution surveys like LAMOST \citep{Luo2015,Bai2021} and DESI \citep{Cooper2023,DESI2024}, with SNR\,$ \sim 5-10$ $\sigma_x \sim 0.1$--$0.2$.}
    \label{fig:noise_effect}
\end{figure*}

\begin{figure*}[htbp]
    \centering
    \includegraphics[width=0.75\textwidth]{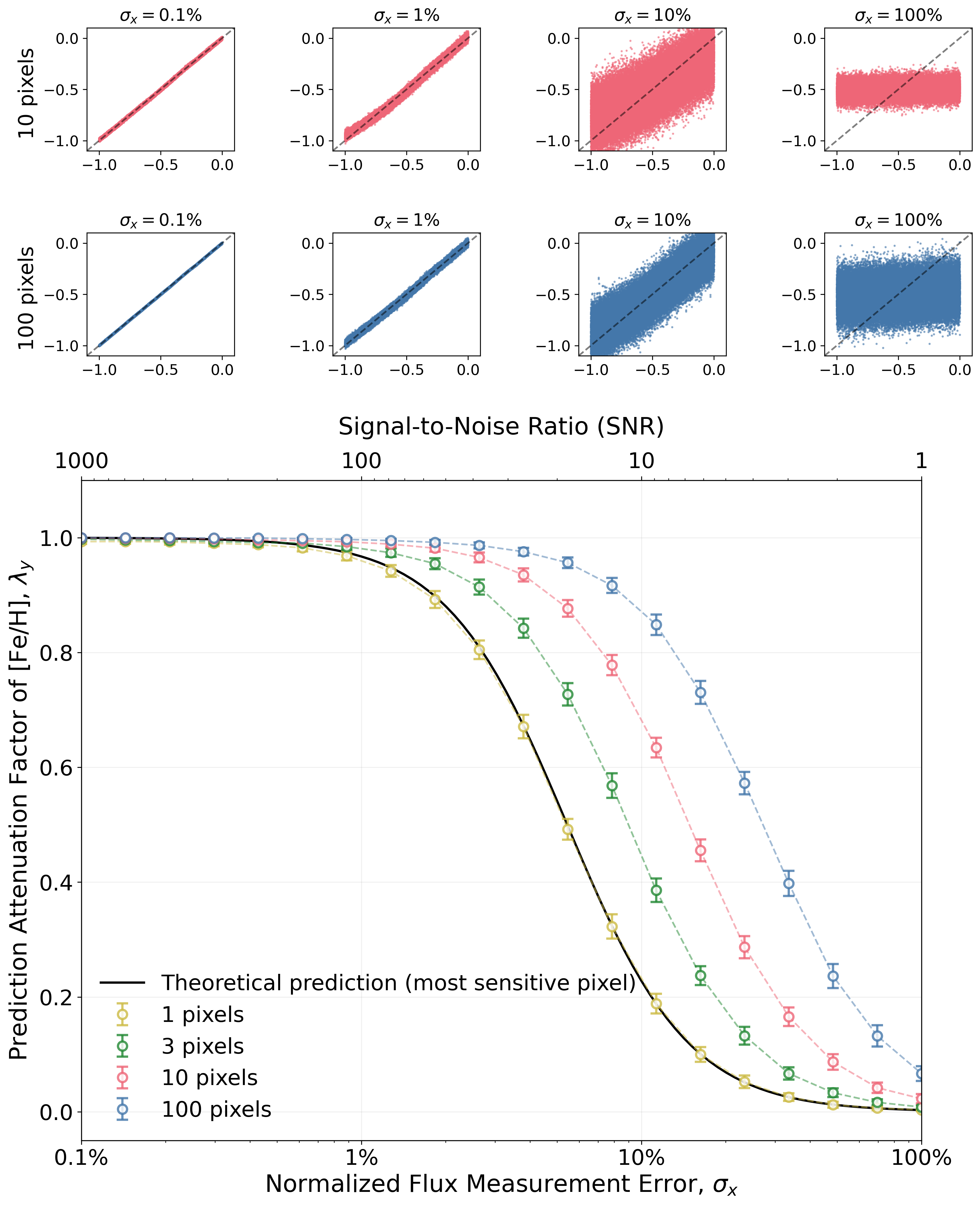}
    \caption{Impact of correlated features on attenuation bias in $\text{[Fe/H]}$ estimation from APOGEE spectra. We select pixels in order of decreasing sensitivity to $\text{[Fe/H]}$. The solid black line shows the theoretical prediction for the most sensitive pixel ($\lambda = 1/(1 + (\sigma_x/\sigma_{\text{range,max}})^2)$), with yellow points showing empirical results using only this pixel. Points show attenuation factors ($\lambda_y$) for different numbers of input features, shown in yellow (1 pixel), green (3 pixels), red (10 pixels), and blue (100 pixels), with error bars indicating $\pm$1 standard deviation across 100 independent trials. Bottom $x$-axis shows measurement uncertainty ($\sigma_x$) in normalized flux units, while top $x$-axis shows the corresponding signal-to-noise ratio. The panels on top show predicted versus true $\text{[Fe/H]}$ for both 10-pixel (top row, red) and 100-pixel (bottom row, blue) cases at selected $\sigma_x$ values. Even at SNR = 100, the 10-pixel case shows visible attenuation, which becomes even more obvious at SNR = 10.}
    \label{fig:attenuation_snr_correlated}
\end{figure*}

Figure \ref{fig:linear_relation} demonstrates both the spectral variations and their relationship with metallicity. The left panel displays synthetic spectra around the Fe absorption line at 16200\AA, illustrating how spectral features vary systematically with $\text{[Fe/H]}$. Throughout this study, we focus on a $\pm$20\AA\ region around this wavelength that contains 179 pixels rich in absorption features, allowing us to study attenuation bias across varying numbers of input dimensions up to $\mathcal{O}(100)$ pixels. To ensure the validity of our linear regression analysis, we restrict our parameter range to a regime where spectral features vary approximately linearly with stellar parameters. Specifically, we fix the effective temperature and surface gravity to typical red clump values ($T_{\text{eff}} = 4750$K, $\log g = 2.5$) and vary only metallicity ($\text{[Fe/H]}$) from $-1.0$ to $0.0$, as this regime is well-tested in the emulation \citep{Ting2019}. For simplicity, we assume all elemental abundances $\text{[X/H]}$ trace $\text{[Fe/H]}$, allowing us to capture the full variation across all pixels.

The right panel examines individual pixels that show more substantial variation ($>0.1$ in normalized flux from $\text{[Fe/H]} = - 1$ to $0$), demonstrating predominantly linear relationships between flux and $\text{[Fe/H]}$. We validate this linearity in the middle panel, where for each pixel, we first fit a linear relation between $\text{[Fe/H]}$ and flux, then predict the flux for 1000 randomly generated spectra from The Payne using their $\text{[Fe/H]}$ values. The residuals from these predictions show that most pixels fall within a 0.1\% range at 1$\sigma$, and even in the worst cases only reach 0.5-1\%, demonstrating narrow confidence intervals around zero across all wavelengths. This confirms that linear relationships provide a sufficient model for our restricted study. Further, recall from Section~\ref{sec:higher-order-regression} that for polynomial terms of order $n$, the attenuation factor scales as $1/(1 + n^2(\sigma_x/\sigma_{\text{range}})^2)$. Since quadratic fits have been shown to better describe normalized flux variations with stellar labels \citep{Rix2016}, and higher order terms suffer from stronger attenuation bias, our analysis focusing on the linear case might be conservative in terms of the true impact of attenuation bias in real spectroscopic applications.

Having established the validity of our linear approximation, we now examine how measurement noise affects spectral inference. Figure \ref{fig:noise_effect} illustrates the progressive degradation of spectral features with increasing noise levels. Starting from a noiseless spectrum ($\sigma_x = 0$), we add uncorrelated Gaussian noise corresponding to different SNR values: SNR = 100 ($\sigma_x = 0.01$), typical of APOGEE and other high-resolution stellar surveys where features remain clearly discernible; SNR = 20 ($\sigma_x = 0.05$), characteristic of low-resolution surveys like LAMOST where noise begins to significantly impact the spectrum; and SNR = 5 ($\sigma_x = 0.2$), common for faint targets where spectral features become severely degraded.

To quantify the attenuation bias in this realistic setting, we generate training and validation sets of 1000 spectra each, with no noise in the $\text{[Fe/H]}$ labels. As established in the theoretical framework, the size of the training set and label precision do not affect the expected attenuation bias. To make our analysis more physically relevant, we sort pixels by their ``sensitivity" - calculated as their standard deviations across the full metallicity range. This approach mimics common practice in spectroscopic analysis where reliable absorption lines of particular atomic species are identified for parameter estimation \citep{Sousa2014,Heiter2021,Smith2021}.

Figure \ref{fig:attenuation_snr_correlated} shows how attenuation bias manifests in this correlated case. The solid black line represents the theoretical prediction for the most sensitive pixel ($\lambda = 1/(1 + (\sigma_x/\sigma_{\text{range,max}})^2)$), which serves as our best-case reference if we have one pixel or if all pixels are independent and equally sensitive. Using only the single most sensitive pixel (yellow points), our simulations closely follow this theoretical prediction, validating our framework in this simplified case where analytical treatment remains tractable.

As expected with correlated features, the empirical results demonstrate a clear departure from the independent features case (see Appendix~\ref{app:independent-mean})—increasing the number of correlated pixels consistently improves the attenuation factor. For 100 pixels at SNR = 100 ($\sigma_x = 1\%$), the attenuation is nearly negligible (mean $\lambda_y = 0.998$), compared to $\lambda_y = 0.6$ seen with independent features. However, this improvement diminishes at lower SNR; even with 100 pixels at SNR = 10, we find $\lambda_y = 0.85$, demonstrating how even correlated features struggle to overcome some of the more reasonable flux measurement noise from modern spectroscopic surveys like LAMOST and DESI.

The practical implications become particularly relevant when considering realistic spectroscopic scenarios. With 10 informative pixels at SNR = 100, we observe $\lambda_y = 0.99$, already introducing a 1\% systematic error. This quickly degrades to $\lambda_y = 0.63$ at SNR = 10. For any reasonable spectral analysis, there are usually only finite features from an atomic species that contribute meaningfully \citep[e.g.][]{Heiter2021,Smith2021}. While major elements like Fe that have myriad of features can achieve minimal attenuation bias by combining $>100$ lines at high SNR, the bias becomes increasingly problematic either as SNR decreases or when analyzing elements with fewer ($<10$) clean spectral features. Indeed, for atomic species that have $\sim$one strong absorption feature (e.g., potassium and vanadium abundances for APOGEE), even at relatively high SNR = 30, we see $\lambda_y = 0.72$, with rapid deterioration at lower SNR. 

The top panels visualize predicted versus true $\text{[Fe/H]}$ relationships at different noise levels for both 10-pixel and 100-pixel cases, where the 10-pixel case shows visible attenuation even at SNR = 100 that becomes severe at SNR = 10, while the 100-pixel case demonstrates better resilience against attenuation through correlated features, although the bias is still quite visible at SNR = 10.

These results demonstrate that while natural correlations in spectral features provide notable protection against attenuation bias, particularly at moderate to high SNR and for labels that contribute to multiple features, this mitigation is not absolute. The effect remains detectable and potentially significant in many practical scenarios encountered in spectroscopic analysis.

\begin{figure*}[htbp]
    \centering
    \includegraphics[width=\textwidth]{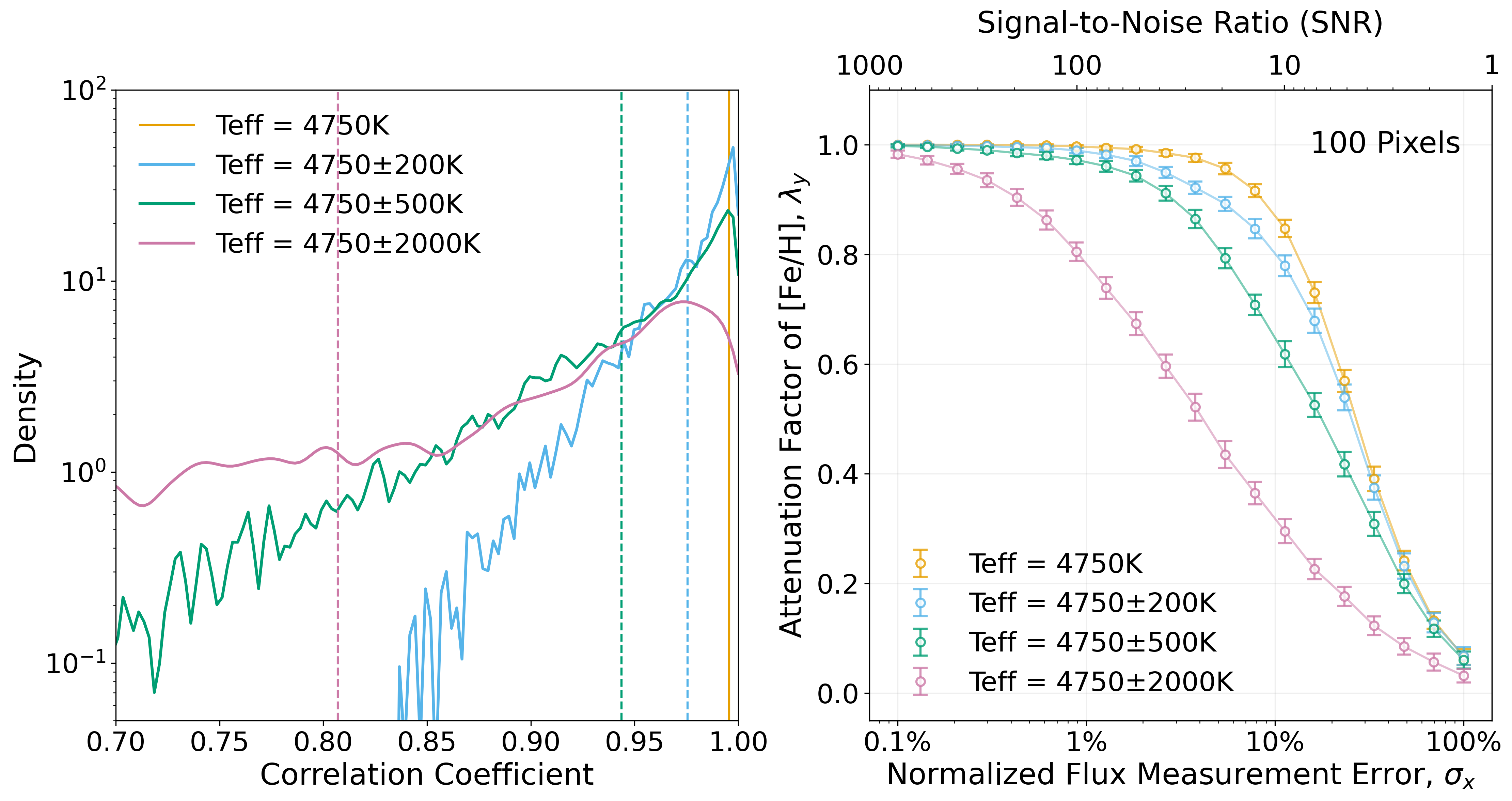}
    \caption{Distribution of pairwise correlation coefficients among the 100 most $\text{[Fe/H]}$-sensitive pixels for different ranges of temperature variation, calculated using kernel density estimation (bandwidth = 0.03). Left panel: Vertical lines indicate mean correlations for each temperature range. Fixed temperature case shown in orange, with temperature variations of $T_{\text{eff}}\pm200$K (blue), $T_{\text{eff}}\pm500$K (green) and $T_{\text{eff}}\pm2000$K (purple). Right panel: Mean attenuation factors and $\pm$1 standard deviation error bars across 100 independent trials for the 100-pixel case. Bottom $x$-axis shows measurement uncertainty ($\sigma_x$) in normalized flux units, top $x$-axis shows corresponding signal-to-noise ratio assuming $\sigma_{\text{range}} = 0.1$.}
    \label{fig:temperature_effect}
\end{figure*}

The analysis above, while being more realistic than the independent feature case, still represents an optimistic scenario. Our assumption of perfect correlation achieved by varying only $\text{[Fe/H]}$ likely overestimates the attenuation mitigation effect compared to real spectra. In practice, multiple stellar labels (especially stellar parameters) vary simultaneously. For example, variations in effective temperature alter the stellar atmosphere, affecting all spectral features—including those sensitive to $\text{[Fe/H]}$—even at fixed metallicity values. This coupling between parameters weakens the perfect correlation between pixels that would arise from $\text{[Fe/H]}$ variations alone.

To show how this idealized scenario breaks down in practice, we examine a simpler case showing how variations in effective temperature affect the correlations between spectral features and their resulting attenuation bias. Figure \ref{fig:temperature_effect} demonstrates this effect by introducing, on top of the $\text{[Fe/H]}$ variation, different levels of temperature scatter around our fiducial red clump temperature ($T_{\text{eff}} = 4750$K) while still fixing the $\log g$. Focusing on the 100-pixel case that showed promising mitigation in our previous analysis, we demonstrate that even in this optimistic high-dimension scenario with high SNR and resolution, adding realistic parameter variations can accentuate again the attenuation bias.

For each temperature range, we calculate the correlation matrix for our 100 most sensitive pixels. Taking the upper triangular portion of this matrix to avoid redundancy, we compute the distribution of unique pairwise correlation coefficients. The left panel shows these distributions using kernel density estimation (KDE with bandwidth 0.03), with vertical lines indicating the mean correlation for each temperature range. By definition with zero temperature spread, we only have correlation of one which is shown as the solid vertical line on right panel. As expected, increasing temperature variation progressively weakens pixel correlations. With no temperature variation (orange), the pixels show nearly perfect correlation as expected from our previous analysis. 

However, as we increase the temperature range to $T_{\text{eff}}\pm200$K (blue), $T_{\text{eff}}\pm500$K (green), and $T_{\text{eff}}\pm2000$K (purple), the correlation distributions broaden and shift toward lower values even for these most sensitive pixels responding to $\text{[Fe/H]}$. The $\pm2000$K range represents an extreme case where temperature information is essentially not measured, allowing us to examine the limiting behavior of decorrelation. This decorrelation occurs because different spectral features respond differently to temperature changes—some features become stronger while others weaken, even at the same $\text{[Fe/H]}$.

This weakened correlation then translates into a smaller mitigation of the attenuation bias as shown in the right panel. The impact on attenuation bias is clear in the right panel. As temperature variations increase and correlations weaken, the attenuation becomes more severe at all SNR levels (here we also assumed $\sigma_{\text{range}} = 0.1$ to calculate the SNR). Even at high SNR = 100, the attenuation factor becomes $\lambda_y = 0.97$ at $T_{\text{eff}}\pm500$K, but then reaches $\lambda_y = 0.81$ at $T_{\text{eff}}\pm2000$K, dropping visibly with increasing temperature spread.

This demonstrates that our earlier analysis with fixed temperature represents an optimistic scenario—in real applications where multiple stellar parameters vary simultaneously corrupting the perfect correlations, the protective effect of correlated features is diminished.

\begin{figure*}[htbp]
    \centering
    \includegraphics[width=\textwidth]{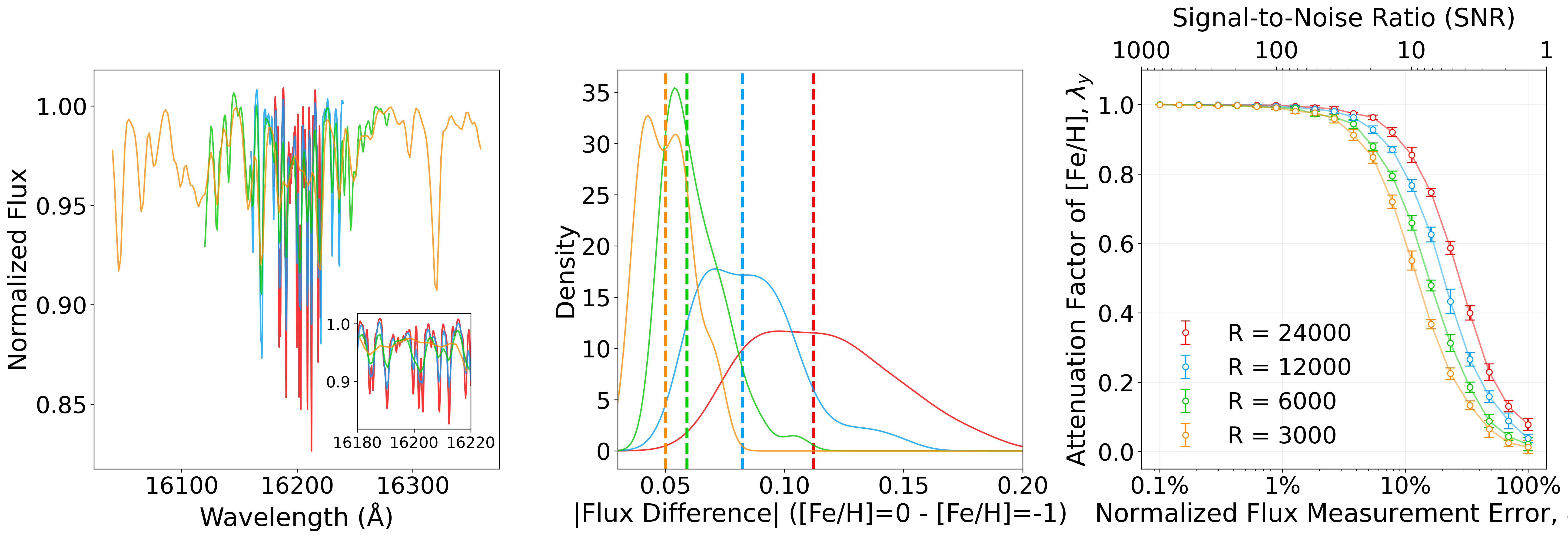}
    \caption{Impact of spectral resolution on feature strength and attenuation bias. Left panel shows example spectra at different resolutions (R = 24000 in red, 12000 in blue, 6000 in green, 3000 in orange), with wavelength range proportionally increased at lower resolutions to maintain the same number of pixels, mimicking fixed CCD real estate. Inset highlights the progressive blending of absorption features with decreasing resolution. Middle panel shows kernel density estimation of absolute flux differences between $\text{[Fe/H]} = -1$ and $0$ spectra for the 50 strongest features at each resolution, with vertical dashed lines indicating median values. Right panel shows the resulting attenuation factors as a function of measurement uncertainty, demonstrating increasingly severe attenuation at lower resolutions. Top axis shows the corresponding SNR assuming the optimistic case of $\sigma_{\text{range}} \sim 0.1$.}
    \label{fig:resolution_effects}
\end{figure*}

Finally, throughout this study, we have focused on APOGEE's high-resolution regime (R = 24,000) as our baseline case. Even in this optimistic scenario, we assumed $\sigma_{\text{range}} \sim 0.1$ based on typical spectral variations. To examine how this assumption and our earlier conclusions might change with spectral resolution, we conduct a systematic analysis across different resolutions from APOGEE (R = 24,000) down to LAMOST/DESI-like resolution (R = 3,000).

To ensure a fair comparison across resolutions, we maintain the same number of pixels by proportionally increasing the wavelength range as resolution decreases, essentially assuming the same CCD real estate. When degrading the spectra, we account for the initial APOGEE resolution by calculating the effective Gaussian kernel width as $\sigma_{\text{eff}} = \sqrt{\sigma_{\text{target}}^2 - \sigma_{\text{APOGEE}}^2}$. 

The left panel of Figure \ref{fig:resolution_effects} illustrates the degradation of spectral features with decreasing resolution. At APOGEE's high resolution ($R=24,000$, red), individual absorption features are clearly resolved, as shown in both the main panel and the zoomed inset. As resolution decreases (blue through orange), these features become progressively blended (see also Figure~\ref{fig:noise_effect}), effectively reducing the information content per pixel as the range of variation ($\sigma_{\text{range}}$) decreases.  For clarity of visualization, we truncate the $R=3,000$ case to the range 16,000\AA-16,400\AA. Following the theoretical framework, this reduction in $\sigma_{\text{range}}$ at fixed SNR should lead to more severe attenuation bias at lower resolutions. While we maintain the same number of pixels across all resolutions, the larger wavelength resolution elements at lower resolution result in broader wavelength coverage.

The middle panel quantifies this degradation by showing the distribution of absolute flux differences between spectra at $\text{[Fe/H]} = -1$ and $0$ for the 50 strongest features at each resolution. The vertical dashed lines indicate the median difference at each resolution. Naively, we should expect the strength to decrease by the ratio of R, but in practice, the low resolution features regain some strength through blending effects.

This analysis refines our earlier assumption of $\sigma_{\text{range}} \sim 0.1$ for APOGEE resolution. While the maximum peak-to-peak variation in flux is indeed about 0.1, the actual $\sigma_{\text{range}}$ is smaller when considering multiple pixels. For example, with 50 pixels, the effective $\sigma_{\text{range}}$ is approximately a factor of 3 smaller (or more precisely, $\sqrt{12}$ smaller assuming uniform distribution). Therefore, while our original assumption of $\sigma_{\text{range}} \sim 0.1$ provides useful intuition, the exact value depends on the number of strongest pixels considered in the analysis.

While we have quantified specific attenuation values, the key insight of this paper is more fundamental: even in optimistic scenarios, attenuation bias remains non-negligible. This conclusion stems from two critical factors: (a) the measurement uncertainty in flux ($\sigma_x$) is typically only a factor of a few smaller than the intrinsic variation range ($\sigma_{\text{range}}$), and (b) although increasing the number of pixels can mitigate attenuation bias when features are correlated, this mitigation is moderate—especially when only ~10 atomic features contribute significantly to the signal, even in the optimistic limit of perfect correlation.

The impact of this feature degradation on attenuation bias is shown in the right panel of Figure~\ref{fig:resolution_effects}. For each resolution, we calculate attenuation factors following the same methodology as our earlier analysis, but now using the resolution-appropriate spectra. The results show progressively more severe attenuation at lower resolutions across all SNR values. Even at relatively high SNR = 100, the attenuation becomes significantly worse as resolution decreases. The systematic degradation of feature strength with decreasing resolution suggests that attenuation bias becomes increasingly important to consider in lower-resolution surveys, even when high SNR can be achieved through long exposures or instrument design. This highlights the fundamental trade-offs between resolution, SNR, and feature strength that must be carefully considered in survey design and analysis.

\section{Discussion and Implications}
\label{sec:discussion}

This study examines the fundamental nature of attenuation bias in astronomical machine learning applications, demonstrating that when directly mapping from input features to output labels—a common practice in machine learning—this bias is inherent and widespread. The challenge stems from how modern machine learning methods, especially neural networks, typically handle uncertainties. While output uncertainties can be incorporated into loss (objective) functions, the treatment of input uncertainties remains challenging within most standard frameworks. This stands in contrast to traditional Bayesian approaches where input uncertainties can be explicitly considered in the analysis.

We find that the magnitude of the attenuation bias depends primarily on input measurement precision, rather than on label precision or training sample size. Using spectral analysis as a case study, we investigate a domain where regression tasks are routinely applied \citep{Ho2017,Guiglion2020,Xiang2019,Angelo2024,Li2024,Zhang2024,Hattori2024}. While we focus on spectroscopy, the implications of attenuation bias extend broadly across observational astronomy, where measurement uncertainties are ubiquitous.

Our analysis centers on linear regression, a choice motivated by two key considerations. First, linear regression permits analytical prediction of attenuation bias under certain assumptions, providing concrete mathematical insights into the effects of measurement uncertainties. Second, despite its simplicity, linear regression serves as a foundation for many astronomical analysis methods. Many power-law relations in astronomy (e.g., Kennicutt-Schmidt relation, Tully-Fisher relation, M-$\sigma$ relation) can be transformed into linear regressions in logarithmic space. Furthermore, many classical machine learning approaches are effectively generalized linear regression with extended, curated features, and neural networks can be viewed as a combination of a feature transformer and linear regression.

\subsection{Why Attenuation Bias Particularly Matters for Astronomy}

While attenuation bias receives relatively little attention in general machine learning applications, it poses a particular challenge for astronomy due to the inherent dynamic range of astronomical measurements. For univariate or independent multivariate cases, the relevant scale is the ratio of signal variance to measurement error ($\sigma_{\text{range}}/\sigma_x$). The theoretical framework predicts that even at a ratio of 10—meaning the dynamic range of the data is approximately tenfold the measurement error—coefficient and label bias manifests at the percent level. While general machine learning applications typically operate with much higher SNR (e.g., natural image processing), astronomy frequently operates in this critical regime where measurement uncertainties are significant relative to the signal range.

This challenge pervades astronomy as we attempt to leverage order-of-magnitude measurements to achieve percent-level understanding through statistical samples. Consider the case study presented here: flux variations in normalized spectra versus flux measurement errors. Even in the optimistic scenario of high-resolution spectra, typical variations are only 5-10\%, while measurement errors in dedicated spectroscopic surveys remain around 1\% \citep{DeSilva2015,Holtzman2018,Gilmore2022}. This places $\sigma_{\text{range}}/\sigma_x$ squarely in the $\mathcal{O}(1)$-$\mathcal{O}(10)$ regime where attenuation bias becomes prominent.

Similar examples appear throughout astronomy. Stellar age measurements, typically have uncertainties of $\mathcal{O}(1)$ Gyr—approximately 10\% of the Hubble time \citep{Labreton2014,Aerts2015,Valle2015,Xiang2017,Wu2018,Valentini2019,Claytor2020,Hayden2022}. Elemental abundance measurements, typically achieve precisions of 0.05-0.1 dex, again about 10\% of the full dynamic range for [X/H] \citep{Feltzing1998,Carretta2002,Kirby2008,Xiang2017,Pancino2010,Holtzman2018}. 

The situation becomes particularly relevant for power-law relationships when transformed into log-log space for linear regression. For instance, supermassive black hole mass measurements spanning three-fiver orders of magnitude can only tolerate uncertainties of about 0.3-0.5 dex \citep[e.g.][]{Merritt2001,Graham2012,Shen2013,Li2023}—corresponding to roughly a factor of three in linear scale—to avoid significant attenuation bias. Similar precision limitations appear in measurements of stellar mass and gas mass, where $\mathcal{O}(0.1)$ dex or more uncertainties are common in multi-dex power law studies \citep{Conroy2009,Swindle2011,Leja2017,Leja2020}. Thus, attenuation bias systematically affects astronomical measurements exactly where precision matters most.

While some studies can tolerate attenuation bias when it merely compresses the dynamic range of labels while preserving relative ordering, this bias becomes crucial when comparing results to physical models or performing parameter estimation. Cosmological studies provide a compelling example: in photometric redshift estimation, high accuracy is essential because biases can propagate through cosmological analyses, significantly affecting parameter inference \citep{Cunha2014,Almosallam2016,McLeod2017,Davis2019,Stolzner2021}. This sensitivity to bias is commonly assessed through the Probability Integral Transform (PIT) test, where a uniformly distributed sample (similar to our study) is simulated to verify recovery of the uniform distribution \citep[e.g.][]{Dey2021,Lin2024}.

Common mitigation strategies in cosmology focus on addressing potential bias sources through careful sample selection—for example, collecting training samples that span the Self-Organizing Map (SOM) space to ensure unbiased representation of galaxy properties, or gathering high-quality spectroscopic redshifts \citep{Masters2015,Wright2020,Johnston2021}. However, these approaches implicitly assume that bias primarily stems from training sample distribution and label quality. Our theoretical framework demonstrates that this assumption may be incomplete, as attenuation bias persists regardless of training sample properties, suggesting that these mitigation strategies warrant reevaluation.

Similar misconceptions appear in other domains, such as the inference of stellar ages from asteroseismic data \citep{ Chaplin2014,Serenelli2017,Li2020}, where calls for collecting more asteroseismic ages reflect the same assumption. Our analysis shows that such efforts, while valuable for other reasons, are unlikely to address the fundamental issue of attenuation bias.

\subsection{Implications of Attenuation Bias for Spectral Analysis with Machine Learning}

While this paper's primary goal is to provide a general framework for understanding attenuation bias—which can be partially predicted in the simplistic limit of linear regression—we showcase a specific application: spectral analysis with machine learning. We chose this example for several compelling reasons. First, as noted above, the ratio between flux measurement error and flux variation naturally falls in the regime where attenuation bias becomes significant. Second, spectra provide an ideal test case for the theoretical framework, offering naturally correlated multivariate features that allow us to extend our analysis beyond univariate cases. Third, spectroscopy represents one of astronomy's most important observational modes, yet remains largely unexplored in general machine learning literature, which has led the astronomical community to independently develop numerous specialized machine learning applications for spectral analysis \citep{Fabbro2018,Leung2019,Ting2019,Guiglion2020}.

The challenge extends beyond traditional spectroscopy to even lower-resolution regimes, including narrow-band filters like the Gaia XP spectra and ground-based surveys such as J-PLUS \citep{Cenarro2019,Wang2022}, S-PLUS \citep{MendesdeOliveira2019}, and JPAS \citep{Bonoli2021,Yuan2023}. Given the persistent challenges in generating robust ab initio spectral models, researchers increasingly favor data-driven approaches, using high-resolution spectral observations or other precise measurements (from eclipsing binaries, astrometry, asteroseismology, or high-quality spectra) as training labels for machine learning models \citep{Suveges2017,Ambrosch2023,Bello-Garcia2023}.

While it is often argued that machine learning models can effectively transfer and match labels across different surveys, our study reveals potential problems with this assumption. Even for elemental abundances under optimal conditions—APOGEE resolution and SNR—attenuation bias remains non-negligible, compressing the dynamic range of inferred labels away from their original scale. Although this bias depends on the number of features employed and the range of stellar parameters (e.g., effective temperature as a nuisance parameter), and strongly correlated features can help minimize the impact, such mitigation is only possible for atomic species with numerous spectral features. Key elements, including widely studied species like oxygen, have limited atomic features in the optical wavelength ranges \citep{Prakapavivcius2013,Ting2018}, leading to potentially severe attenuation (e.g. fig 13 in \citealt{Xiang2019}, fig 4 in \citealt{Zhang2024}). This effect is frequently observed in label transfer applications, particularly at low resolution, where one-to-one comparisons between training labels and predictions show more prominent skewing for elements with weaker and/or fewer features \citep[e.g., figure 9-11][]{Leung2019}. 

This attenuation bias can have important implications, particularly for fundamental measurements of Galactic structure. For example, spectra contain information about stellar parameters that, when combined with apparent magnitudes, enable estimation of absolute magnitudes and hence distances—a variation of spectroscopic-photometric distance determination. The systematic underestimation of distances due to attenuation bias, whic to our knowledge has not been well recognized or treated, can have cascading effects on our understanding of Galactic dynamics and chemical evolution.

When these attenuated distances are used to map the Milky Way's rotation curve, the radius $R$ at which the orbital velocity $v_c$ is measured is systematically underestimated. Given that the enclosed mass within radius $R$ is determined by $M(<R) = v_c^2 R/G$, the percentage attenuation in distance measurements directly translates to the same percentage underestimation in the enclosed mass $M(<R)$. As we have established in this study, attenuation bias in low-resolution surveys can be significant even under conservative estimates, reaching the percent level. This systematic underestimation of the Milky Way's mass has far-reaching implications for our understanding of the Galactic rotation curve, dark matter distribution, and overall Galactic dynamics. These findings suggest the need to revisit data-driven spectroscopic-photometric distance determinations in studies of  the Milky Way.

\subsection{Mitigation Strategies for Attenuation Bias}

Given the significant implications of attenuation bias, it is natural to consider potential mitigation strategies. The most obvious approach involves calibration, either theoretical or empirical, acknowledging that this fundamental bias stems naturally from input uncertainty and is independent of training sample size and label accuracy.

Yet, theoretical calibration through analytic calculations has its own limitations. While in this study we have shown that bias can be accurately predicted for univariate cases with homogeneous errors when uncertainties are well-measured, real data typically exhibits heteroskedastic (inhomogeneous) errors, making theoretical calculations substantially more complex.

More generally, as demonstrated in our multivariate linear regression analysis, theoretical prediction becomes challenging when features are not completely independent. Even in the linear case with homogeneous errors, the theoretical framework requires impractical assumptions such as perfect correlation between features and stringent constraints on linear coefficients. These theoretical limitations become even more pronounced for nonlinear machine learning methods with unknown causality structures between the variables.

Empirical calibration of attenuation bias presents significant challenges. In the simplest case where both input $x$ and output $y$ have homogeneous errors, one might assume that the attenuation bias $\lambda_y$ applies uniformly to all data points and could be calibrated globally using validation data. However, in practice, both inputs and outputs typically exhibit heteroskedastic errors, leading to different attenuation factors for individual data points. Consequently, global calibration, while potentially mitigating the overall bias, likely introduces new systematic effects that are difficult to control.

Simulation-based calibration might seem like an alternative approach, as demonstrated in our spectral analysis. However, the need for data-driven machine learning often arises precisely because physical models contain non-negligible systematics. As we have shown, attenuation bias strongly depends on the statistical properties of correlated features (spectral features in our case), making simulation-based calibration potentially unreliable when these properties are not well understood. Nonetheless, our spectral analysis suggests a possible path forward: for a given set of underlying physical models, the attenuation bias scales predictably with SNR, potentially enabling data-driven empirical calibration based on the empirical SNR dependence of the attenuation bias.

Our study has shown that when features are not independent, their combined effect can help negate random input noise, thereby mitigating attenuation bias. This suggests that when possible, using data in its more raw form with multiple correlated estimators—each subject to independent noise—can be beneficial. For example, full spectral fitting based on individual pixels could help mitigate these issues compared to approaches using limited spectral indices or summary statistics that restrict the number of input features. Furthermore, more restricted models like linear regression might offer advantages in that their bias can be better predicted or calibrated, and they suffer less from attenuation compared to super-linear regression. As demonstrated in this study, linear regression permits analytical prediction of attenuation bias under certain assumptions, whereas for super-linear regression, such theoretical predictions become intractable.

More broadly, while our study focuses primarily on error-ignorant discriminative models—which unfortunately remain common in astronomical applications, especially with modern architectures like neural networks—the community has developed several sophisticated statistical approaches to address measurement uncertainties. Errors-in-variables models explicitly incorporate measurement error variances in the input to provide unbiased parameter estimates. Bayesian hierarchical models offer another powerful framework, particularly well-suited to handling complex error structures and incorporating prior information. These approaches can account for measurement uncertainties at multiple levels, from individual observations to population-level parameters, providing a more robust foundation for astronomical inference.

\subsection{Discriminative Models versus Generative Models}

Our analysis highlights an asymmetry in how measurement errors affect regression models: uncertainties in dependent variables (outputs) do not cause attenuation bias, while uncertainties in independent variables (inputs) systematically bias the results. This insight has important implications for choosing between discriminative and generative approaches in astronomical data analysis.

Discriminative models, which directly map from observables to physical parameters (e.g., from spectra to stellar properties), treat the noisy measurements as input features. As we have demonstrated through our theoretical framework, measurement uncertainties in these inputs inevitably lead to attenuation bias in an error-ignorant framework, regardless of training sample size or label quality. 

In contrast, generative models first learn to predict observables from physical parameters, then use optimization methods (e.g., $\chi^2$ minimization) or sampling approaches (e.g., MCMC) to infer parameters from data. This approach naturally accommodates measurement uncertainties in two ways. First, during the forward modeling phase, measurement uncertainties appear in the dependent variables where they only affect variance of the model estimate, not bias. Second, during the parameter inference phase, measurement uncertainties can be explicitly incorporated into the likelihood function or objective function.

This theoretical advantage becomes particularly relevant in scenarios where training labels have higher precision than the observed data. Consider, for example, the common task of analyzing low-resolution survey spectra using training labels derived from high-resolution observations. Discriminative models must contend with significant measurement uncertainties in their input features (the low-resolution spectra), leading to inevitable attenuation bias. Generative models, by learning the forward mapping from parameters to spectra, can better preserve the precision of the training labels.

\section{Conclusion}
\label{sec:conclusion}

A persistent challenge in astronomical machine learning has been systematic bias where predictions compress the dynamic range of true values---the highest true values are consistently predicted too low while the lowest values are predicted too high. This bias is particularly relevant in modern astronomy, where data-driven models increasingly attempt to extract information from low-quality data by leveraging high-quality training labels from other companion observations. While conventional wisdom has attributed this bias to factors such as inaccurate labels, insufficient training data, or non-uniform sampling of training data, the theoretical analysis reveals a more fundamental origin, and surprisingly demonstrates that all of these commonly cited factors are irrelevant to solving this issue.

Through examination of linear regression with noisy data, we demonstrate that this systematic bias naturally emerges from attenuation bias (or regression dilution)---a direct consequence of how measurement uncertainties in input variables spread the observed distribution and fundamentally lead to underestimation of regression coefficients. This bias persists regardless of training sample size, label accuracy, or parameter distribution. 

For univariate or independent multivariate cases, we prove that the attenuation factor $\lambda$ follows a simple relationship. This relationship reveals that even with signal-variance-to-noise ratios of 10 ($\sigma_{\text{range}}/\sigma_x = 10$), predictions suffer from 1\% bias, with the effect becoming severe at lower ratio. Since this range of signal-variance-to-noise of 10 and lower is common in astronomical observations but rare in typical machine learning applications, this effect has been largely overlooked in the broader astronomy literature despite its significant impact on astronomical inference.

Our further investigation of multivariate linear regression yields several insights: for independent features, the bias depends solely on the ratio of measurement uncertainty to signal variance, independent of sample size or dimensionality. Adding more independent features cannot overcome this fundamental limitation. However, we demonstrate that correlated features can help mitigate this bias. The degree of mitigation depends on both the number of correlated features and their correlation strength, which is particularly relevant for spectroscopic analysis where features are naturally correlated through underlying physics. In the theoretical limit of infinite perfectly correlated features, the attenuation bias can be completely resolved, though this limit is rarely achievable in practice.

Using high-resolution stellar spectroscopy as a case study, we show that even in optimal conditions---APOGEE resolution ($R = 24,000$) with high SNR---attenuation bias remains detectable. The effect becomes substantially more severe at lower resolutions or SNR values typical of surveys like LAMOST and DESI. This has particular implications for abundance measurements of elements with limited spectral features, where the protective effect of correlated features is reduced. Furthermore, we demonstrate that variations in other stellar parameters (like effective temperature) can weaken feature correlations, diminishing their ability to mitigate attenuation bias. 

The attenuation biases of 1-10\% we have identified can have profound implications for fundamental astronomical relationships, particularly when propagated through subsequent analyses. The naive application of machine learning models that ignore input uncertainties can introduce systematic biases in key parameter estimations. These biases could significantly impact our understanding of fundamental astronomical relationships across multiple scales: from galactic-scale relations like the Kennicutt-Schmidt law and the M-$\sigma$ relation between black hole and galaxy properties, to more specific spectroscopic analyses such as age-metallicity correlations and the Milky Way's rotation curve.

This attenuation bias stems from the fundamental asymmetry in how measurement errors affect regression models: uncertainties in dependent variables (outputs) only increase variance, while uncertainties in independent variables (inputs) directly contribute to attenuation bias. This asymmetry has important implications for model choice in astronomical applications. Generative models that predict observables (such as spectra) from physical parameters, followed by parameter inference through optimization or sampling, may be practically more robust than discriminative approaches that attempt to infer parameters directly from noisy observations. This advantage arises because in generative models, the noisier measurements appear in the dependent variables where they only affect variance, while in discriminative approaches, these same noisy measurements serve as independent variables where they inevitably introduce attenuation bias.

While our analytical treatment focuses on linear regression and spectral analysis, the insights about attenuation bias naturally extend to more complex nonlinear models as well as all applications of machine learning in the astronomical context.  As astronomy continues to embrace larger surveys with varying data quality, understanding and accounting for attenuation bias becomes increasingly crucial for reaching more robust and accurate inference, ensuring that upstream label determination does not inadvertently propagate systematic errors to downstream scientific conclusions.

\vspace{0.5cm}

I am grateful to my colleagues at OSU whose lively discussions during our daily astro-coffee sessions have inspired some of this work. Special thanks to David Weinberg for his careful and invaluable feedback as this paper took shape. The proximity of our offices provided both intellectual stimulation and emotional urgency to complete this work. Some ideas in this research emerged from my ritual of bouncing thoughts off o1-preview and claude-3.5-sonnet - while their initial suggestions often led me down fascinating (and largely unproductive) rabbit holes, our back-and-forth exchanges proved invaluable. Alongside my human colleagues at OSU, I'd like to thank o1-preview and claude for their unwavering intellectual companionship during the course of this research. This research is supported by the National Science Foundation under Grant No. AST-2406729.


\bibliography{manuscriptNotes.bib}
\bibliographystyle{aasjournal}

\appendix

\section{Attenuation Bias Under Generalized Least Squares (GLS)}
\label{app:gls}

In the main text, we established that measurement errors in the independent variable ($x$) lead to systematic underestimation of regression coefficients, known as attenuation bias, when using ordinary least squares (OLS) regression. A natural question arises: can employing Generalized Least Squares (GLS) mitigate this bias? GLS can provide more efficient estimates than OLS by accounting for known heteroskedasticity or correlation structures in the residuals of the dependent variable. However, as we will demonstrate, the fundamental attenuation bias due to errors in $x$ remains unaltered by the switch from OLS to GLS.

Consider the univariate linear model:
\begin{equation}
y_{\text{true},i} = \beta x_{\text{true},i},
\end{equation}
with observed values given by:
\begin{align}
x_{\text{obs},i} &= x_{\text{true},i} + \delta_{x,i}, \\
y_{\text{obs},i} &= y_{\text{true},i} + \delta_{y,i},
\end{align}
where $\mathbb{E}[\delta_{x,i}] = 0$, $\mathbb{E}[\delta_{y,i}] = 0$, $\text{Var}(\delta_{x,i}) = \sigma_x^2$, and $\delta_{x,i}$, $\delta_{y,i}$ are independent of $x_{\text{true},i}$. The variance of the measurement errors in $y$ may vary with $i$, so we define $w_i = 1/\sigma_{y,i}^2$ as the GLS weight.

The GLS estimator for the slope $\beta$ takes the form:
\begin{equation}
\hat{\beta}_{GLS} = \frac{\sum_i w_i x_{\text{obs},i} y_{\text{obs},i}}{\sum_i w_i x_{\text{obs},i}^2}.
\end{equation}
Substituting $y_{\text{obs},i} = \beta x_{\text{true},i} + \delta_{y,i}$ and $x_{\text{obs},i} = x_{\text{true},i} + \delta_{x,i}$:
\begin{equation}
\hat{\beta}_{GLS} = \frac{\sum_i w_i (x_{\text{true},i} + \delta_{x,i})(\beta x_{\text{true},i} + \delta_{y,i})}{\sum_i w_i (x_{\text{true},i} + \delta_{x,i})^2}.
\end{equation}

Taking the expectation and using $\mathbb{E}[\delta_{x,i}] = \mathbb{E}[\delta_{y,i}] = 0$, along with the independence assumptions, we obtain:
\begin{align}
\mathbb{E}[\hat{\beta}_{GLS}] &= \frac{\sum_i w_i \mathbb{E}[x_{\text{true},i}\beta x_{\text{true},i}]}{\sum_i w_i (\sigma_{\text{range}}^2 + \sigma_x^2)} \nonumber \\
&= \frac{\sum_i w_i \beta \sigma_{\text{range}}^2}{\sum_i w_i (\sigma_{\text{range}}^2 + \sigma_x^2)}.
\end{align}

Since the weights $w_i$ are just multiplicative factors, they cancel out in the proportional sense:
\begin{equation}
\mathbb{E}[\hat{\beta}_{GLS}] = \beta \frac{\sigma_{\text{range}}^2}{\sigma_{\text{range}}^2 + \sigma_x^2}.
\end{equation}

This result is exactly the same attenuation factor derived for OLS. Thus, even if the error model in $y$ is accounted for by GLS, the fundamental bias from measurement errors in $x$ persists. The weighting of observations by $w_i$ does not alter the ratio that characterizes attenuation: it only affects the relative contribution of each data point, not the underlying relationship between $x_{\text{obs}}$ and $y_{\text{obs}}$.

To demonstrate, we repeat the simulations described in the main text (see Section~\ref{sec:theory-univariate}), but now using GLS instead of OLS. The simulation setup remains largely the same, with a two-dimensional parameter space defined by the ratio of signal range to measurement error ($\sigma_{\text{range}}/\sigma_x$) and the measurement uncertainty in the dependent variable ($\sigma_y$). The key difference lies in the regression method: we now employ GLS, with weights set to $1/\sigma_y^2$ to account for the known variances in $y$.

The GLS estimates for the coefficients $\hat{\beta} = [\hat{\beta}_1, \hat{\beta}_0]$ are obtained using the normal equation:
$$\hat{\beta} = (X^\top W X)^{-1} X^\top W y$$
where $X$ is the design matrix (observed $x$ values with a column of ones for the intercept) and $W$ is the diagonal weight matrix. We then compute the predicted $y$ values using these estimated coefficients and perform a simple OLS regression of the predicted values against the true values to obtain the attenuation factor $\lambda_y$ (the slope of this regression).

Our numerical simulations further confirm that the attenuation factor $\lambda_y$ depends only on $\sigma_{\text{range}}/\sigma_x$ and is independent of $\sigma_y$, even when using GLS regression. This demonstrates that GLS does not mitigate the fundamental attenuation bias introduced by measurement errors in the independent variable $x$.

\section{Variance of Linear Coefficient Estimator}
\label{app:variance}

While the attenuation bias in the linear coefficient estimator is independent of both the training set size $n$ and the uncertainty in the dependent variable $\sigma_y$, the variance of the estimator depends on both quantities. Here we derive this dependence explicitly for centered variables (i.e., $\overline{x}_{\text{obs}} = \overline{y}_{\text{obs}} = 0$).

To derive the variance of $\hat{\beta}$ in the univariate case, let's begin with its fundamental expression and systematically show how it relates to the residuals. Starting with the OLS estimator:
\begin{equation}
    \hat{\beta} = \frac{\sum_i x_{\text{obs},i} y_{\text{obs},i}}{\sum_i x_{\text{obs},i}^2}
\end{equation}
We can substitute $y_{\text{obs},i} = \beta x_{\text{true},i} + \delta_{y,i} = \beta x_{\text{obs},i} - \beta \delta_{x,i} + \delta_{y,i}$ into this expression:
\begin{equation}
    \hat{\beta} = \frac{\sum_i x_{\text{obs},i}(\beta x_{\text{obs},i} - \beta \delta_{x,i} + \delta_{y,i})}{\sum_i x_{\text{obs},i}^2}
\end{equation}
Distributing $x_{\text{obs},i}$ gives us:
\begin{equation}
    \hat{\beta} = \beta + \frac{\sum_i x_{\text{obs},i}(-\beta \delta_{x,i} + \delta_{y,i})}{\sum_i x_{\text{obs},i}^2}
\end{equation}

Now, to find $\operatorname{Var}(\hat{\beta})$, we note that $\beta$ is constant, so:
\begin{equation}
    \operatorname{Var}(\hat{\beta}) = \operatorname{Var}\left(\frac{\sum_i x_{\text{obs},i}(-\beta \delta_{x,i} + \delta_{y,i})}{\sum_i x_{\text{obs},i}^2}\right)
\end{equation}
To evaluate this variance, treating $x_{\text{obs},i}/\sum_i x_{\text{obs},i}^2$ as our constant (conditional on $x_{\text{obs}}$) and applying this property element-wise:
\begin{align}
    \operatorname{Var}(\hat{\beta}) &= \sum_i \left(\frac{x_{\text{obs},i}}{\sum_j x_{\text{obs},j}^2}\right)^2 \operatorname{Var}(-\beta \delta_{x,i} + \delta_{y,i}) \nonumber \\
    &= \frac{\sum_i x_{\text{obs},i}^2}{(\sum_i x_{\text{obs},i}^2)^2} (\sigma_y^2 + \beta^2 \sigma_x^2)
\end{align}
For large samples, we can use the fact that $\frac{1}{n}\sum_i x_{\text{obs},i}^2$ approaches $\operatorname{Var}(x_{\text{obs}}) = \sigma_{\text{range}}^2 + \sigma_x^2$. Therefore:
\begin{equation}
    \operatorname{Var}(\hat{\beta}) = \frac{\sigma_y^2 + \beta^2 \sigma_x^2}{n(\sigma_{\text{range}}^2 + \sigma_x^2)}.
\end{equation}

This result reveals several important insights about the precision of our coefficient estimates. First, unlike the attenuation bias discussed in the main text, the variance scales inversely with sample size $n$, indicating that larger training sets do improve estimation precision even if they cannot address the fundamental bias. Second, measurement uncertainties in both $x$ and $y$ contribute to the variance through the numerator term, showing how both sources of error affect our estimation uncertainty.

Having derived the variance of the linear coefficient estimator in the univariate case, we now extend this analysis to the multivariate setting with $p$ features. We maintain our previous assumptions of centered variables ($\overline{\mathbf{x}} = \mathbf{0}$, $\overline{y}=0$) and homogeneous uncertainties ($\sigma_x$ for all features), while adding the requirement that features are independent and identically distributed with variance $\sigma_{\text{range}}^2$.

In the multivariate setting, the ordinary least squares estimator for each coefficient takes the form:
\begin{equation}
    \hat{\beta}_j = \frac{\sum_i x_{\text{obs},ij}(\sum_{k=1}^p \beta_k x_{\text{true},ik} + \delta_{y,i})}{\sum_i x_{\text{obs},ij}^2}
\end{equation}

To analyze this estimator, we first substitute $x_{\text{obs},ij} = x_{\text{true},ij} + \delta_{x,ij}$ in the appropriate terms:
\begin{align}
    \hat{\beta}_j &= \frac{\sum_i x_{\text{obs},ij}(\sum_{k=1}^p \beta_k x_{\text{true},ik} + \delta_{y,i})}{\sum_i x_{\text{obs},ij}^2} \nonumber \\
    &= \frac{\sum_i x_{\text{obs},ij}(\beta_j x_{\text{true},ij} + \sum_{k\neq j} \beta_k x_{\text{true},ik} + \delta_{y,i})}{\sum_i x_{\text{obs},ij}^2} \nonumber \\
    &= \frac{\sum_i x_{\text{obs},ij}(\beta_j(x_{\text{obs},ij} - \delta_{x,ij}) + \sum_{k\neq j} \beta_k x_{\text{true},ik} + \delta_{y,i})}{\sum_i x_{\text{obs},ij}^2} \nonumber \\
    &= \frac{\sum_i x_{\text{obs},ij}(\beta_j x_{\text{obs},ij} - \beta_j \delta_{x,ij} + \sum_{k\neq j} \beta_k x_{\text{true},ik} + \delta_{y,i})}{\sum_i x_{\text{obs},ij}^2} \nonumber \\
    &= \beta_j + \frac{\sum_i x_{\text{obs},ij}(-\beta_j \delta_{x,ij} + \delta_{y,i} + \sum_{k\neq j} \beta_k x_{\text{true},ik})}{\sum_i x_{\text{obs},ij}^2}
\end{align}

To compute $\operatorname{Var}(\hat{\beta}_j)$, we condition on the observed features $x_{\text{obs}}$, treating $x_{\text{obs},ij}/\sum_i x_{\text{obs},ij}^2$ as constant, similar to our approach in the univariate case:
\begin{align}
    \operatorname{Var}(\hat{\beta}_j) &= \sum_i \left(\frac{x_{\text{obs},ij}}{\sum_\ell x_{\text{obs},\ell j}^2}\right)^2 \operatorname{Var}(-\beta_j \delta_{x,ij} + \delta_{y,i} + \sum_{k\neq j} \beta_k x_{\text{true},ik})
\end{align}

The independence of $\delta_{x,ij}$, $\delta_{y,i}$, and $x_{\text{true},ik}$ allows us to decompose the variance term:
\begin{align}
    \operatorname{Var}(-\beta_j \delta_{x,ij} + \delta_{y,i} + \sum_{k\neq j} \beta_k x_{\text{true},ik}) &= \beta_j^2\sigma_x^2 + \sigma_y^2 + \sum_{k\neq j} \beta_k^2\operatorname{Var}(x_{\text{true},ik})
\end{align}

The key step comes in evaluating $\operatorname{Var}(x_{\text{true},ik})$ conditional on $x_{\text{obs}}$:
\begin{align}
    \operatorname{Var}(x_{\text{true},ik} | x_{\text{obs}}) &= \operatorname{Var}(x_{\text{obs},ik} - \delta_{x,ik} | x_{\text{obs}}) \nonumber \\
    &= \operatorname{Var}(-\delta_{x,ik} | x_{\text{obs}}) \nonumber \\
    &= \sigma_x^2
\end{align}

Substituting back and simplifying:
\begin{align}
    \operatorname{Var}(\hat{\beta}_j) &= \frac{\sum_i x_{\text{obs},ij}^2}{(\sum_i x_{\text{obs},ij}^2)^2} (\sigma_y^2 + \sum_{k=1}^p \beta_k^2 \sigma_x^2)
\end{align}

For large samples, $\frac{1}{n}\sum_i x_{\text{obs},ij}^2$ approaches $\sigma_{\text{range}}^2 + \sigma_x^2$ for each feature $j$, yielding our final result:
\begin{equation}
    \operatorname{Var}(\hat{\beta}_j) = \frac{\sigma_y^2 + \sum_{k=1}^p \beta_k^2 \sigma_x^2}{n(\sigma_{\text{range}}^2 + \sigma_x^2)}
\end{equation}

This multivariate generalization reveals how the variance of each coefficient estimator depends not only on its own parameter value but also on all other coefficients in the model. The denominator retains the same form as in the univariate case, showing that the fundamental scaling with sample size and feature variance remains unchanged. However, the numerator now includes contributions from all features' measurement uncertainties, weighted by their respective coefficients, reflecting how noise in any feature affects the precision of all coefficient estimates. As the dimension $p$ increases, the variance of individual estimators grows through the summation term $\sum_{k=1}^p \beta_k^2 \sigma_x^2$. This behavior is not surprising: given the same constraints from the training data, the determination of the hyperplane becomes more stochastic at higher dimensions, reflecting the fundamental challenge of maintaining precision in high-dimensional parameter estimation.

\section{Derivation of the attenuation bias for correlated multivariate features}
\label{app:multivariate}

We have shown in Section~\ref{sec:multivariate-theory-indepedent} that in the case of independent features, the number of dimensions in the input would not mitigate the attenuation bias, and the attenuation bias calculation follows the case of the univariate scenario. However, when the features are correlated, the random noise would need to conspire to create the attenuation bias, making it possible to reduce the attenuation bias as the dimensionality increases.

In general, deriving an analytic theory for multivariate linear regression with correlated features is challenging. For simplicity, the results we have laid out in Section~\ref{sec:multivariate-correlated-theory} assume perfect correlation between all the features. In this case, without loss of generality, we can assume $x_{\text{true},j} = a_j x_{\text{true}}$, where $a_j$ is a scaling factor for the $j$-th feature, with $j$ ranging from 1 to $p$, and $x_{\text{true}}$ is a scalar random variable representing the underlying true signal.

Recall that, from Equation~\ref{eq:expectation-beta-multivariate}, we have:
\begin{equation}
\mathbb{E}[\hat{\boldsymbol{\beta}}] \approx (\mathbb{E}[\mathbf{X}_{\text{true}}^\top \mathbf{X}_{\text{true}}] + n\sigma_x^2 \mathbf{I}_p)^{-1} \mathbb{E}[\mathbf{X}_{\text{true}}^\top \mathbf{X}_{\text{true}} ]\boldsymbol{\beta}
\label{eq:expectation_approximation}
\end{equation}

We will calculate the individual terms in this expression. First, we compute $\mathbb{E}[\mathbf{X}_{\text{true}}^\top \mathbf{X}_{\text{true}}]$. Under the assumption of perfect correlation, $\mathbf{X}_{\text{true}}$ can be expressed as:
\begin{equation}
\mathbf{X}_{\text{true}} = x_{\text{true}} \mathbf{A},
\end{equation}
where $x_{\text{true}}$ is the random variable representing the underlying intrinsic input data, and $\mathbf{A}$ is a fixed $1 \times p$ row vector of scaling factors $[a_1, a_2, \dots, a_p]$.

Now, we can calculate $\mathbb{E}[\mathbf{X}_{\text{true}}^\top \mathbf{X}_{\text{true}}]$:
\begin{align}
\mathbb{E}[\mathbf{X}_{\text{true}}^\top \mathbf{X}_{\text{true}}] &= n \, \mathbb{E}[(x_{\text{true}} \mathbf{A})^\top (x_{\text{true}} \mathbf{A})] \nonumber \\
&= n\, \mathbb{E}[x_{\text{true}}^2] \mathbf{A}^\top \mathbf{A}.
\end{align}
Let $S_x = \sum_{i=1}^n x_{\text{true},i}^2 = n\, \mathbb{E}[x_{\text{true}}^2]$, which represents the evaluation of the expectation  over $n$ samples of the training data. Thus, the matrix $\mathbb{E}[\mathbf{X}_{\text{true}}^\top \mathbf{X}_{\text{true}}]$ can be estimated as:
\begin{equation}
\mathbb{E}[\mathbf{X}_{\text{true}}^\top \mathbf{X}_{\text{true}}] \approx  S_x \mathbf{A}^\top \mathbf{A}.
\end{equation}
Substituting these expressions into Equation \eqref{eq:expectation_approximation}, we have:
\begin{align}
\mathbb{E}[ \hat{\boldsymbol{\beta}} ] &\approx \left( S_x \mathbf{A}^\top \mathbf{A} + n \sigma_x^2 \mathbf{I}_p \right)^{-1} S_x \mathbf{A}^\top \mathbf{A} \boldsymbol{\beta}. \nonumber \\
&= \left( \mathbf{A}^\top \mathbf{A} + \frac{n \sigma_x^2}{S_{x}} \mathbf{I}_p \right)^{-1} \mathbf{A}^\top \mathbf{A} \boldsymbol{\beta}.
\label{eq:expected_beta_scaled}
\end{align}

Let $\kappa = \frac{n \sigma_x^2}{S_{x}}$, which represents the relative magnitude of the measurement error variance to the signal variance scaled by the sample size. To compute $(\mathbf{A}^\top \mathbf{A} + \kappa \mathbf{I}_p)^{-1}$, note that $\mathbf{A}^\top \mathbf{A}$ is a rank-1 matrix if $\mathbf{A}$ is a non-zero vector. Let $\mathbf{a}$ be the $p \times 1$ vector of scaling factors $a_j$. Then:
\begin{equation}
\mathbf{A}^\top \mathbf{A} = \mathbf{a} \mathbf{a}^\top.
\end{equation}
Therefore, the matrix we need to invert is:
\begin{equation}
\mathbf{M} = \mathbf{a} \mathbf{a}^\top + \kappa \mathbf{I}_p.
\end{equation}

We can use the Sherman-Morrison formula to find the inverse of $\mathbf{M}$:
\begin{equation}
( \mathbf{a} \mathbf{a}^\top + \kappa \mathbf{I}_p )^{-1} = \kappa^{-1} \left( \mathbf{I}_p - \frac{ \mathbf{a} \mathbf{a}^\top }{ \kappa + \mathbf{a}^\top \mathbf{a} } \right).
\end{equation}
Substituting back, we have:
\begin{align}
\mathbb{E}[ \hat{\boldsymbol{\beta}} ] &\approx \kappa^{-1} \left( \mathbf{I}_p - \frac{ \mathbf{a} \mathbf{a}^\top }{ \kappa + \mathbf{a}^\top \mathbf{a} } \right) \mathbf{a} \mathbf{a}^\top \boldsymbol{\beta} \nonumber \\
&= \kappa^{-1} \left( \mathbf{a} \mathbf{a}^\top \boldsymbol{\beta} - \frac{ \mathbf{a} \mathbf{a}^\top \mathbf{a} \mathbf{a}^\top }{ \kappa + \mathbf{a}^\top \mathbf{a} } \boldsymbol{\beta} \right).
\end{align}
Note that $\mathbf{a}^\top \boldsymbol{\beta}$ is a scalar, so $\mathbf{a} \mathbf{a}^\top \boldsymbol{\beta} = (\mathbf{a}^\top \boldsymbol{\beta})\mathbf{a}$. Similarly, $\mathbf{a}^\top \mathbf{a} = \sum_{j=1}^p a_j^2$. Compute the numerator in the second term:
\begin{equation}
\mathbf{a} \mathbf{a}^\top \mathbf{a} = ( \mathbf{a}^\top \mathbf{a} ) \mathbf{a}.
\end{equation}
Therefore, the expression simplifies to:
\begin{align}
\mathbb{E}[ \hat{\boldsymbol{\beta}} ] &\approx \kappa^{-1} \left( ( \mathbf{a}^\top \boldsymbol{\beta} ) \mathbf{a} - \frac{ ( \mathbf{a}^\top \boldsymbol{\beta} )( \mathbf{a}^\top \mathbf{a} ) \mathbf{a} }{ \kappa + \mathbf{a}^\top \mathbf{a} } \right) \nonumber \\
&= \kappa^{-1} ( \mathbf{a}^\top \boldsymbol{\beta} ) \mathbf{a} \left( 1 - \frac{ \mathbf{a}^\top \mathbf{a} }{ \kappa + \mathbf{a}^\top \mathbf{a} } \right).
\end{align}
Simplify the term in parentheses:
\begin{align}
1 - \frac{ \mathbf{a}^\top \mathbf{a} }{ \kappa + \mathbf{a}^\top \mathbf{a} } &= \frac{ \kappa }{ \kappa + \mathbf{a}^\top \mathbf{a} }.
\end{align}
Therefore:
\begin{align}
\mathbb{E}[ \hat{\boldsymbol{\beta}} ] &\approx \kappa^{-1} ( \mathbf{a}^\top \boldsymbol{\beta} ) \mathbf{a} \left( \frac{ \kappa }{ \kappa + \mathbf{a}^\top \mathbf{a} } \right ) \nonumber \\
&= \frac{ \mathbf{a}^\top \boldsymbol{\beta} }{ \kappa + \mathbf{a}^\top \mathbf{a} } \mathbf{a} \\
&\equiv \lambda_\beta \mathbf{a}.
\end{align}
We have the expected attenuation factor:
\begin{equation}
\lambda_\beta = \frac{ \mathbf{a}^\top \boldsymbol{\beta} }{ \kappa + \mathbf{a}^\top \mathbf{a} }.
\end{equation}

However, in this form, we still cannot write $\mathbb{E}[ \hat{\boldsymbol{\beta}} ]$ as a factor of the true $\boldsymbol{\beta}$. In general, this is not possible. However, if we further assume that $\boldsymbol{\beta} = \beta \mathbf{a}$, where $\beta$ is a scalar coefficient, then $\mathbf{a}^\top \boldsymbol{\beta} = \beta \mathbf{a}^\top \mathbf{a}$, and the expected value of $\hat{\boldsymbol{\beta}}$ becomes:
\begin{equation}
\mathbb{E}[ \hat{\boldsymbol{\beta}} ] \approx \frac{ \beta \mathbf{a}^\top \mathbf{a} }{ \kappa + \mathbf{a}^\top \mathbf{a} } \mathbf{a}.
\end{equation}
Thus, each element of $\mathbb{E}[ \hat{\boldsymbol{\beta}} ]$ is:
\begin{equation}
\mathbb{E}[ \hat{\beta}_j ] \approx \lambda_\beta \beta a_j = \lambda_\beta \beta_j,
\label{eq:beta_j_expected_scaled}
\end{equation}
and we arrive at the analytic expression for the attenuation factor $\lambda_\beta$:
\begin{equation}
\lambda_\beta = \frac{ \mathbf{a}^\top \mathbf{a} }{ \kappa + \mathbf{a}^\top \mathbf{a} } = \frac{ \sum_{j=1}^{p} a_j^2 }{ \frac{ n \sigma_x^2 }{ S_{x} } + \sum_{j=1}^{p} a_j^2 }.
\end{equation}
This is the expression we discussed in Section~\ref{sec:multivariate-correlated-theory}. We derived this analytic formula under strong assumptions about the scalability and alignment between $\boldsymbol{\beta}$ and $\mathbf{a}$. While such explicit expressions may not exist for general multivariate linear regression, the goal of this calculation is to demonstrate a key principle: as the number of non-trivial and correlated features increases, the attenuation bias decreases, with the attenuation factor approaching unity.

\begin{figure*}[htbp]
    \centering
    \includegraphics[width=\textwidth]{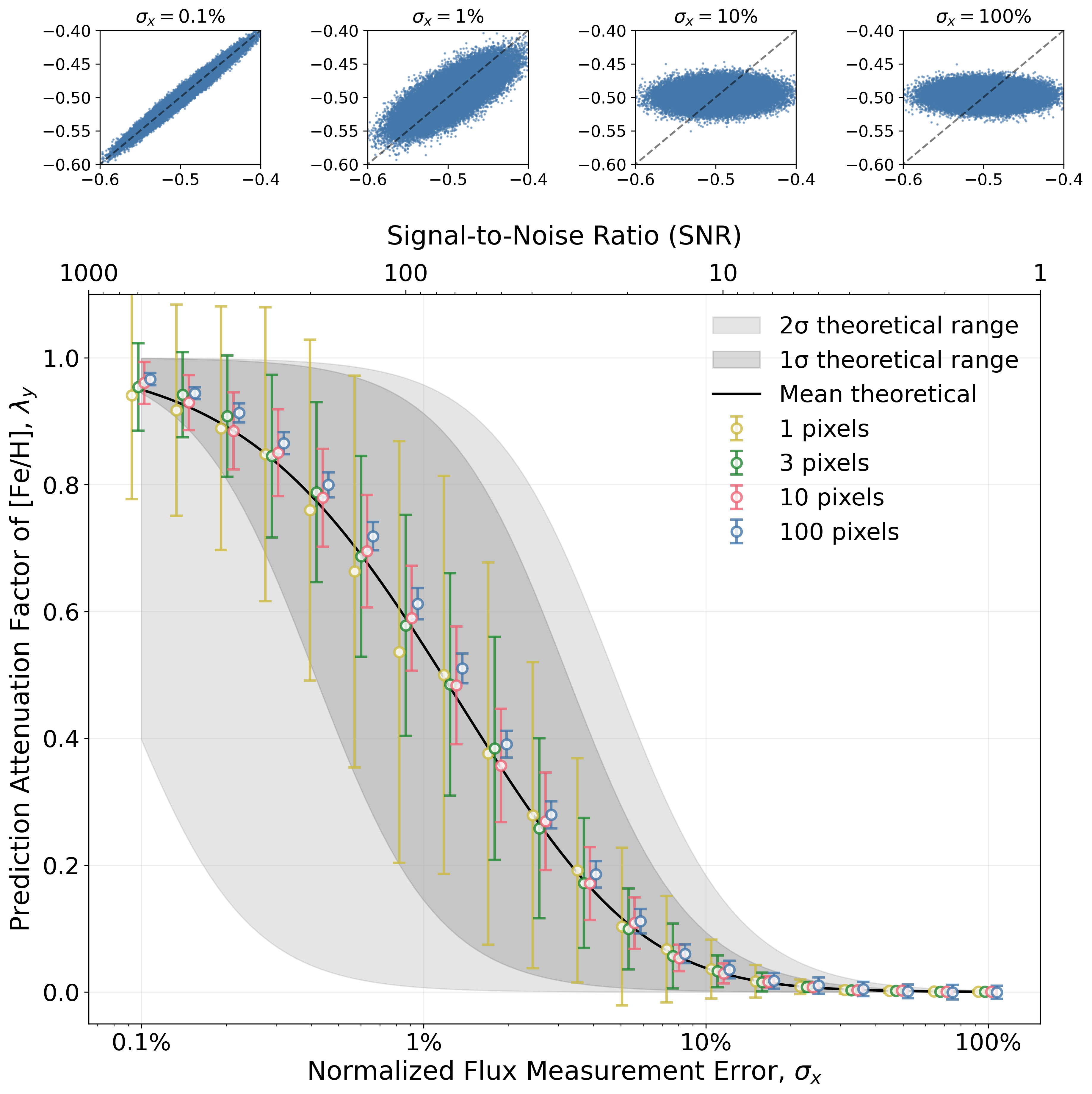}
    \caption{Empirical validation of theoretical predictions for attenuation bias with independent features of varying strengths. Using synthetic APOGEE spectra where pixel-wise correlations are artificially broken. Gray shaded regions show theoretical 1$\sigma$ (dark) and 2$\sigma$ (light) ranges of attenuation factors, reflecting the spread in feature sensitivities across different pixels. Colored points show empirical attenuation factors ($\lambda_y$) for different numbers of input features ($p = 1, 3, 10, 100$ pixels in yellow, green, red, and blue respectively), with error bars indicating $\pm$1 standard deviation across 100 independent trials. Bottom $x$-axis shows measurement uncertainty ($\sigma_x$) in normalized flux units, with corresponding signal-to-noise ratio on top. Top panels show predicted versus true $\text{[Fe/H]}$ for the 100-pixel case at selected $\sigma_x$ values.}
    \label{fig:attenuation_snr}
\end{figure*}

\section{Empirical Validation of Independent Feature Predictions with Varying Feature Strengths}
\label{app:independent-mean}

In Section~\ref{sec:multivariate-theory-indepedent}, we demonstrated that for independent features with equal strengths, attenuation bias persists regardless of the number of features used. For features with varying sensitivities, the theoretical framework predicts that the effective attenuation factor represents a weighted average of individual feature attenuations, with the final predictions being pulled toward the mean of the parameter range. Here, we use synthetic APOGEE spectra to validate these theoretical predictions.

In spectroscopy, features are naturally correlated through underlying physics—when varying a single parameter like $\text{[Fe/H]}$, all spectral features should respond in a physically consistent way. To test the theoretical predictions about independent features, we need to artificially break these physical correlations. We begin by generating 1,000 synthetic spectra using The Payne spectral emulator, focusing on the same $\pm$20\AA\ wavelength range around 16200\AA\ as in our main analysis, with $\text{[Fe/H]}$ uniformly distributed from $-1$ to $0$. For each pixel, we derive an empirical linear relationship between flux and $\text{[Fe/H]}$ from these synthetic spectra. As validated in Figure \ref{fig:linear_relation}, these linear relationships adequately describe the pixel-wise dependence of flux on metallicity.

Using these empirical relationships, we create a pool of 10,000 samples by independently sampling flux values at each pixel, maintaining the observed flux distributions but breaking the natural correlations between pixels. For each sample, we estimate $\text{[Fe/H]}$ by inverting these linear relationships pixel-by-pixel, then take the mean across all pixels as the final $\text{[Fe/H]}$ label. This approach, by treating pixels independently, allows different $\text{[Fe/H]}$ values to contribute to the final label from each pixel. Because we treat all pixels as contributing independently to $\text{[Fe/H]}$, the final label average close to the mean of the input range ($\text{[Fe/H]} \in [-0.6,-0.4]$).

Using multivariate linear regression, we analyze the relationship between the label $\text{[Fe/H]}$ ($y$) and the normalized flux values ($\mathbf{x}$) across wavelength bins, with each pixel $j$ having a corresponding coefficient $\beta_j$. For each simulation configuration, we train the model using 1,000 mock spectra with known $\text{[Fe/H]}$ values and validate on another 1,000 independent mock spectra.

Figure \ref{fig:attenuation_snr} demonstrates how the attenuation factor varies with measurement uncertainty and number of input pixels. The gray shaded regions show the theoretical predictions for attenuation factors expected across all 179 pixels, with darker and lighter bands indicating 1$\sigma$ (16th-84th percentile) and 2$\sigma$ (2nd-98th percentile) ranges respectively. This spread arises because different pixels show varying sensitivities to $\text{[Fe/H]}$ (different $\sigma_{\text{range}}$), leading to different attenuation factors at fixed $\sigma_x$.

To validate these predictions, we perform 100 independent trials for each configuration of pixel number ($p = 1, 3, 10, 100$) and measurement uncertainty ($\sigma_x$). The colored points show the mean attenuation factor ($\lambda_y$) from these trials, with error bars indicating the standard deviation. These results empirically validate the theoretical predictions about independent features with varying strengths. When features contribute independently and have different sensitivities ($\sigma_{\text{range}}$), they experience different degrees of attenuation ($\lambda_{\beta,j}$), and their weighted average pulls the final prediction toward the mean.

Also as predicted by the theoretical framework for independent features, the mean attenuation shows no significant dependence on the number of pixels used—demonstrating that including more independent features cannot mitigate the bias. The variance does decrease as more pixels are included, reflected in smaller error bars for larger $p$, because with fewer pixels, each random selection might include predominantly strong or weak features, leading to more variable attenuation.

The top panels show predicted versus true $\text{[Fe/H]}$ for the 100-pixel case at different noise levels, illustrating increasing attenuation as measurement uncertainty grows. Notably, even at relatively high SNR = 100 ($\sigma_x = 0.01$), we observe measurable attenuation. 

We note that this represents a pessimistic limit of independent features. In real physical systems, where features are naturally correlated, increasing the number of features can help mitigate the attenuation bias—a point we explored in detail in the main text.

\end{CJK*}
\end{document}